\pdfoutput=1
\documentclass[11pt,twoside,a4paper,cmspaper,final,collab]{cms-tdr}
\def\svnVersion{2eef2e8}\def\svnDate{2023/09/09}\def\cmsCernNoTag{CERN-EP-2022-125}\def\cmsCernDate{\today}\def\cmsMessage{Published in the Journal of High Energy Physics as \href{http://dx.doi.org/10.1007/JHEP09(2023)032}{\doi{10.1007/JHEP09(2023)032}.}}
\begin{document}\cmsNoteHeader{HIG-21-010}

\providecommand{\cmsTopLeft}{top left\xspace}
\providecommand{\cmsBottom}{bottom\xspace}
\providecommand{\cmsTopRight}{top right\xspace}
\providecommand{\cmsLeft}{left\xspace}
\providecommand{\cmsRight}{right\xspace}
\providecommand{\cmsTable}[1]{\resizebox{\textwidth}{!}{#1}}

\newlength\cmsTabSkip\setlength{\cmsTabSkip}{2ex}
\newcommand{\orderOf}[1]{\ensuremath{\mathcal{O}(#1)}}
\newcommand{\intLumi}{138\fbinv}
\newcommand{\massLow}{300}
\newcommand{\massHigh}{700}
\newcommand{\mSignal}{\ensuremath{\mHpm=500\GeV}}
\newcommand{\abseta}{\ensuremath{\abs{\eta}}\xspace}
\newcommand{\mtop}{\ensuremath{m_{\PQt}}\xspace}
\newcommand{\mresTop}{\ensuremath{m_{\rTop}}\xspace}
\newcommand{\mbottom}{\ensuremath{m_{\PQb}}\xspace}
\newcommand{\mh}{\ensuremath{m_{\Ph}}\xspace}
\newcommand{\mH}{\ensuremath{m_{\PH}}\xspace}
\newcommand{\mHpm}{\ensuremath{m_{\PHpm}}\xspace}
\newcommand{\mHpmLight}{\ensuremath{\smash[b]{\mHpm<\mtop-\mbottom}}\xspace}
\newcommand{\mHpmIntermediate}{\ensuremath{\smash[b]{\mHpm\approx\mtop}}\xspace}
\newcommand{\mHpmHeavy}{\ensuremath{\smash[b]{\mHpm>\mtop+\mbottom}}\xspace}
\newcommand{\massRange}{\massLow\ to \massHigh\GeV}
\newcommand{\ppToHplustb}{\ensuremath{\Pp\Pp \to \PQt (\PQb) \PH^{+}}\xspace}
\newcommand{\HToTT}{\ensuremath{\PH\to\PGt\PGt}\xspace}
\newcommand{\HpmToTauNu}{\ensuremath{\PHpm \to \PGt^{\pm} \PGnGt}\xspace}
\newcommand{\HpmToCS}{\ensuremath{\PHpm \to \PQc \PQs}\xspace}
\newcommand{\HpmToCB}{\ensuremath{\PHpm \to \PQc\PQb}\xspace}
\newcommand{\HpmToTB}{\ensuremath{\PHpm \to \PQt\PQb}\xspace}
\newcommand{\HpmTohW}{\ensuremath{\PHpm \to \Ph \PWpm}\xspace}
\newcommand{\HpmToHW}{\ensuremath{\PHpm \to \PH \PWpm}\xspace}
\newcommand{\HpToHW}{\ensuremath{\PH^{+} \to \PH \PWp}\xspace}
\newcommand{\HpmToWZ}{\ensuremath{\PHpm \to \PWpm \PZ}\xspace}
\newcommand{\HpmToWA}{\ensuremath{\PHpm \to \PWpm \PSA}\xspace}
\newcommand{\BToSGamma}{\ensuremath{\PQb\to\PQs\gamma}\xspace}
\newcommand{\xsDef}{\ensuremath{\sigma_{\PHpm}\BRto{\HpmToHW, \HToTT}}\xspace}
\newcommand{\xsLow}{0.085}
\newcommand{\xsHigh}{0.019}
\newcommand{\xsLimit}{\xsLow\unit{pb} at \massLow\GeV to \xsHigh\unit{pb} at \massHigh\GeV\xspace}
\newcommand{\refeq}[1]{Eq.~(\ref{eq:#1})\xspace}
\newcommand{\refsec}[1]{Section~\ref{sec:#1}\xspace}
\newcommand{\reftrisec}[3]{Sections~\ref{sec:#1}, \ref{sec:#2}, and \ref{sec:#3}\xspace}
\newcommand{\reffigplural}{Figs.}
\newcommand{\reffig}[1]{Fig.~\ref{fig:#1}\xspace}
\newcommand{\refdifig}[2]{\reffigplural~\ref{fig:#1} and~\ref{fig:#2}\xspace}
\newcommand{\reftab}[1]{Table~\ref{tab:#1}\xspace}
\newcommand{\refcite}[1]{Ref.~\cite{#1}\xspace}
\newcommand{\refcites}[1]{Refs.~\cite{#1}\xspace}
\newcommand{\bjet}{{\PQb} jet\xspace}
\newcommand{\bjetAdj}{{\PQb}-jet\xspace}
\newcommand{\bjets}{{\PQb} jets\xspace}
\newcommand{\rTop}{\ensuremath{\PQt^{\text{res}}}\xspace}
\newcommand{\rIso}[1]{\ensuremath{I_{\text{rel}}^{#1}}\xspace}
\newcommand{\etaphi}{\ensuremath{\eta\text{-}\phi}\xspace}
\newcommand{\tauFR}{\ensuremath{\mathcal{R}_{\JetToTauh}}\xspace}
\newcommand{\tauFRindex}[1]{\ensuremath{\mathcal{R}_{\JetToTauh,~#1}}\xspace}
\newcommand{\sumBins}[1]{\ensuremath{\sum\limits_{#1}}\xspace}
\newcommand{\yDeepTau}[1]{\ensuremath{y}_{#1}\xspace}
\newcommand{\yDeepTauDef}{\ensuremath{\yDeepTau{\alpha}}\xspace}
\newcommand{\dDeepTau}[1]{\ensuremath{D_{#1}}\xspace}
\newcommand{\dR}[2]{\ensuremath{\Delta R(#1, #2)}\xspace}
\newcommand{\dPhi}[2]{\ensuremath{\Delta \phi(#1, #2)}\xspace}
\newcommand{\charge}[1]{\ensuremath{Q_{#1}}\xspace}
\newcommand{\etau}{\ensuremath{\Pe\tauh}\xspace}
\newcommand{\mtau}{\ensuremath{\PGm\tauh}\xspace}
\newcommand{\tauhtauh}{\ensuremath{\tauh\tauh}\xspace}
\newcommand{\ltauhORtauhtauh}{\ensuremath{\star\tauh}\xspace}
\newcommand{\ltauh}{\ensuremath{\ell\tauh}\xspace}
\newcommand{\etautau}{\ensuremath{\Pe\tauh\tauh}\xspace}
\newcommand{\mtautau}{\ensuremath{\PGm\tauh\tauh}\xspace}
\newcommand{\myFS}{\etau, \mtau, \etautau, and \mtautau}
\newcommand{\genuineTauh}{genuine \tauh}
\newcommand{\LToTauh}{\ensuremath{\ell \to \tauh}\xspace}
\newcommand{\EToTauh}{\ensuremath{\Pe \to \tauh}\xspace}
\newcommand{\MToTauh}{\ensuremath{\PGm \to \tauh}\xspace}
\newcommand{\JetToTauh}{\ensuremath{j \to \tauh}\xspace}
\newcommand{\fakeTauh}{misidentified \tauh}
\newcommand{\ltau}{\ensuremath{\ell\tauh}\xspace}
\newcommand{\ltautau}{\ensuremath{\ell\tauh\tauh}\xspace}
\newcommand{\SingleTop}{single \PQt}
\newcommand{\DibosonS}{$\PW\PZ$, $\PZ\PZ$, $\PW\PW$\xspace}
\newcommand{\Vjets}{\PV{+}jets\xspace}
\newcommand{\VjetsDef}{$\PV=\PZ$ or $\PW$}
\newcommand{\ttX}{\ensuremath{\PQt\PAQt\PX}\xspace}
\newcommand{\tanbeta}{\ensuremath{\tan\beta}\xspace}
\newcommand{\sinBetaMinusAlpha}{\ensuremath{\sin(\beta-\mixingAngle)}\xspace}
\newcommand{\alignmentLimit}{\ensuremath{\sinBetaMinusAlpha=1}\xspace}
\newcommand{\nonAlignmentLimit}{\ensuremath{\sinBetaMinusAlpha\neq1}\xspace}
\newcommand{\mixingAngle}{\ensuremath{\alpha}\xspace}
\newcommand{\BRto}[1]{\ensuremath{\mathcal{B}(#1)}\xspace}
\newcommand{\LT}{\ensuremath{L_{\mathrm{T}}}\xspace}
\newcommand{\ST}{\ensuremath{S_{\mathrm{T}}}\xspace}
\newcommand{\SumTwoJetOverHT}{\ensuremath{\frac{\pt^{j_{1}j_{2}}}{\HT}}\xspace}
\newcommand{\PtjjPtHAsym}{\ensuremath{\frac{\pt^{j_{1}j_{2}} - \pt^{\PHpm}}{\pt^{j_{1}j_{2}} + \pt^{\PHpm}}}\xspace}
\newcommand{\ThirdJetOverHToneline}{\ensuremath{{\pt^{j_{3}}/{\HT}}}\xspace}
\newcommand{\ThirdJetOverHT}{\ensuremath{\frac{\pt^{j_{3}}}{\HT}}\xspace}
\newcommand{\SumTwoJetAndLTOverHT}{\ensuremath{\frac{\pt^{j_{1}j_{2}} + \LT}{\HT}}\xspace}
\newcommand{\UES}{unclustered energy scale\xspace}
\newcommand{\TES}{\tauh energy scale\xspace}
\newcommand{\JES}{jet energy scale\xspace}
\newcommand{\JER}{jet energy resolution\xspace}
\newcommand{\mtMuTau}{\ensuremath{\mT(\PGm, \tauh, j_{1}, j_{2}, \ptvecmiss)}\xspace}
\newcommand{\mtLtau}{\ensuremath{\mT(\ell, \tauh, j_{1}, j_{2}, \ptvecmiss)}\xspace}
\newcommand{\invMass}[2]{\ensuremath{m(#1,\, #2)}\xspace}
\newcommand{\UE}{underlying event\xspace}
\newcommand{\hdamp}{\ensuremath{h_{\text{damp}}}\xspace}

\newif\ifRF\RFfalse
\newcommand{\RF}{\ifRF RF\else renormalization and factorization (RF)\RFtrue\fi\xspace}

\newif\ifPythia\Pythiafalse
\newcommand{\PYTHIAeight}{\ifPythia \PYTHIA~v8.212\else \PYTHIA~v8.212~\cite{Sjostrand:2014zea}~\Pythiatrue\fi\xspace}

\newif\ifPHtwo\PHtwofalse
\newcommand{\POWHEGtwo}{\ifPHtwo \POWHEG~v2.0\else \POWHEG~v2.0~\cite{Nason:2004rx, Frixione:2007vw, Alioli:2010xd, Alioli:2010xa,Alioli:2008tz, Bagnaschi:2011tu}\PHtwotrue\fi\xspace}

\newif\ifSR\SRfalse
\newcommand{\SR}{\ifSR SR\else signal region (SR)\SRtrue\fi\xspace}
\newcommand{\SRs}{\ifSR SRs\else signal regions (SRs)\SRtrue\fi\xspace}

\newif\ifVR\VRfalse
\newcommand{\VR}{\ifVR VR\else validation region (VR)\VRtrue\fi\xspace}
\newcommand{\VRs}{\ifVR VRs\else validation regions (VRs)\VRtrue\fi\xspace}

\newif\ifCR\CRfalse
\newcommand{\CR}{\ifCR CR\else control region (CR)\CRtrue\fi\xspace}
\newcommand{\CRs}{\ifCR CRs\else control regions (CRs)\CRtrue\fi\xspace}

\newif\ifMLM\MLMfalse
\newcommand{\MLM}{\ifMLM MLM\else MLM\MLMtrue\fi\xspace}

\newif\ifTT\TTfalse
\newcommand{\SMttbar}{\ifTT \ttbar\else top quark-antiquark pair\xspace\TTtrue\fi}

\newif\ifLO\LOfalse
\newcommand{\LO}{\ifLO LO\else leading-order (LO)\LOtrue\fi\xspace}

\newif\ifNLO\NLOfalse
\newcommand{\NLO}{\ifNLO NLO\else next-to-LO (NLO)\NLOtrue\fi\xspace}

\newif\ifNNLO\NNLOfalse
\newcommand{\NNLO}{\ifNNLO NNLO\else next-to-NLO (NNLO)\NNLOtrue\fi\xspace}

\newif\ifOS\OSfalse
\newcommand{\oSign}{\ifOS OS\else opposite-sign (OS)\OStrue\fi\xspace}

\newif\ifSS\SSfalse
\newcommand{\sSign}{\ifSS SS\else same-sign (SS)\SStrue\fi\xspace}

\newif\ifMC\MCfalse
\newcommand{\MC}{\ifMC MC\else Monte Carlo (MC)\MCtrue\fi\xspace}

\newif\ifSM\SMfalse
\newcommand{\SM}{\ifSM SM\else standard model (SM)\SMtrue\fi\xspace}

\newcommand{\BEH}{Brout--Englert--Higgs\xspace}

\newif\ifBR\BRfalse
\newcommand{\BRs}{\ifBR $\mathcal{B}$\else branching fractions $\mathcal{B}$\BRtrue\fi\xspace}

\newif\ifPDF\PDFfalse
\newcommand{\PDF}{\ifPDF PDF\else parton distribution function (PDF)\PDFtrue\fi\xspace}
\newcommand{\PDFs}{\ifPDF PDFs\else parton distribution functions (PDFs)\PDFtrue\fi\xspace}

\newif\ifME\MEfalse
\newcommand{\ME}{\ifME ME\else matrix element (ME)\MEtrue\fi\xspace}

\newif\ifHD\HDfalse
\newcommand{\TwoHDM}{\ifHD 2HDM\else two-Higgs-doublet model (2HDM)~\cite{Gunion:2002zf,Akeroyd:2016ymd,Branco:2011iw,Craig:2012vn}\HDtrue\fi\xspace}
\newcommand{\TwoHDMs}{\ifHD 2HDMs\else two-Higgs-doublet models (2HDMs)~\cite{Gunion:2002zf,Akeroyd:2016ymd,Branco:2011iw,Craig:2012vn}\HDtrue\fi\xspace}

\newif\ifDiboson\Dibosonfalse
\newcommand{\Diboson}{\ifDiboson diboson\else diboson (\DibosonS)\Dibosontrue\fi\xspace}

\newif\ifEW\EWfalse
\newcommand{\EW}{\ifEW electroweak\else electroweak processes including \Vjets with \VjetsDef and \Diboson production\EWtrue\fi\xspace}

\newif\ifTAUJET\TAUJETfalse
\newcommand{\taujet}{\ifTAUJET \tauh\else tau lepton decaying hadronically (\tauh)\xspace\TAUJETtrue\fi}

\newif\ifBDTG\BDTGfalse
\newcommand{\BDTG}{\ifBDTG BDTG\else boosted decision tree with gradient boost (BDTG)\BDTGtrue\fi\xspace}

\newif\ifQCD\QCDfalse
\newcommand{\QCD}{\ifQCD QCD\else quantum chromodynamics (QCD)\QCDtrue\fi\xspace}

\newif\ifRESTOP\RESTOPfalse
\newcommand{\resTop}{\ifRESTOP \rTop\else resolved top quark (\rTop)\xspace\RESTOPtrue\fi}

\newif\ifLEP\LEPfalse
\newcommand{\leptons}{\ifLEP leptons\else leptons  ($\ell=\Pe,\,\PGm$)\LEPtrue\fi\xspace}
\newcommand{\lepton}{\ifLEP lepton\else lepton ($\ell=\Pe,\,\PGm$)\LEPtrue\fi\xspace}

\newif\ifPV\PVfalse
\newcommand{\PVtx}{\ifPV PV\else primary \pp interaction vertex (PV)\PVtrue\fi\xspace}

\newif\ifPP\PPfalse
\newcommand{\pp}{\ifPP $\Pp\Pp$\else proton-proton ($\Pp\Pp$)\PPtrue\fi\xspace}

\newif\ifMVA\MVAfalse
\newcommand{\MVA}{\ifMVA MVA\else multivariate analysis (MVA)\MVAtrue\fi\xspace}

\newif\ifPF\PFfalse
\newcommand{\PF}{\ifPF PF\else particle-flow (PF)\PFtrue\fi\xspace}

\newif\ifHLT\HLTfalse
\newcommand{\HLT}{\ifHLT HLT\else high-level trigger (HLT)\HLTtrue\fi\xspace}

\newif\ifECAL\ECALfalse
\newcommand{\ECAL}{\ifECAL ECAL\else electromagnetic calorimeter (ECAL)\ECALtrue\fi\xspace}

\newif\ifHCAL\HCALfalse
\newcommand{\HCAL}{hadron calorimeter\xspace}

\newif\ifMADSPIN\MADSPINfalse
\newcommand{\MADSPIN}{\ifMADSPIN \textsc{madspin}\else \textsc{madspin}~\cite{Artoisenet:2012st}\MADSPINtrue\fi\xspace}

\newif\ifDEEPJET\DEEPJETfalse
\newcommand{\DEEPJET}{\ifDEEPJET \textsc{DeepJet}\else \textsc{DeepJet}\DEEPJETtrue\fi\xspace}

\newif\ifDEEPTAU\DEEPTAUfalse
\newcommand{\DEEPTAU}{\ifDEEPTAU \textsc{DeepTau}\else \textsc{DeepTau}~\cite{TAU-20-001,CMS-DP-2019-033}\DEEPTAUtrue\fi\xspace}

\newif\ifTMVA\TMVAfalse
\newcommand{\TMVA}{\ifTMVA \textsc{TMVA}\else \textsc{TMVA}~\cite{Hocker:2007ht}\TMVAtrue\fi\xspace}

\cmsNoteHeader{HIG-21-010}
\title{Search for a charged Higgs boson decaying into a heavy neutral Higgs boson and a \texorpdfstring{$\PW$}{W} boson in proton-proton collisions at \texorpdfstring{$\sqrt{s}=13\TeV$}{sqrt(s) = 13 TeV}}

\date{\today}

\abstract{
  A search for a charged Higgs boson \PHpm decaying
  into a heavy neutral Higgs boson \PH and a \PW boson
  is presented. The analysis targets the \PH decay into a pair
  of tau leptons with at least one of them decaying hadronically and
  with an additional electron or muon present in the event.
  The search is based on proton-proton collision data
  recorded by the CMS experiment during 2016--2018 at
  $\sqrt{s} = 13\TeV$, corresponding to an integrated
  luminosity of 138\fbinv. The data are consistent with
  standard model background expectations. Upper limits at 95\% confidence
  level are set on the product of the cross section and branching fraction
  for an \PHpm in the mass range of 300--700\GeV, assuming an \PH
  with a mass of 200\GeV. The observed limits range from
  0.085\unit{pb} for an \PHpm mass of
  300\GeV to 0.019\unit{pb} for a mass of
  700\GeV. These are the first limits on \PHpm
  production in the $\PHpm \to \PH \PWpm$ decay channel at the LHC.
}

\hypersetup{
  pdfauthor={CMS Collaboration},
  pdftitle    = {Search for a charged Higgs boson decaying into a heavy neutral Higgs boson and a W boson in proton-proton collisions at sqrt(s) = 13 TeV},
  pdfsubject  = {CMS},
  pdfkeywords = {CMS, Higgs boson, charged Higgs boson, 2HDM, MSSM, BSM}
}

\maketitle

\section{Introduction}
\label{sec:introduction}

{\tolerance=800
The experimental confirmation of the \BEH
mechanism~\cite{Higgs:1964ia,Higgs:1964pj,Guralnik:1964eu,Englert:1964et,Higgs:1966ev,Kibble:1967sv}
at the CERN LHC~\cite{Aad:2012tfa,Chatrchyan:2012xdj,Chatrchyan:2013lba}
has provided the long-sought solution to the electroweak symmetry
breaking problem. It has also further established the \SM of particle
physics as a successful theory. Subsequent precision measurements of
the couplings of the observed Higgs boson with the \SM particles are
in agreement with expectations, with an experimental precision of
$\approx$5--33\%~\cite{Khachatryan:2016vau,Sirunyan:2018koj,Aad:2019mbh,Sirunyan:2019twz}.
Regardless of the success it has achieved, the \SM is still considered
to be an effective field theory valid only at low energies because of
its inability to address various fundamental theoretical problems and
compelling observations in nature such as the naturalness problem, the vacuum
metastability, the conjectured cosmological inflation, the
presence of dark matter and the matter-antimatter asymmetry of our universe.
\par}

Numerous theoretical models have been proposed to remedy the
shortcomings of the \SM, many of which predict that
the Higgs sector must also be extended.
Minimal extensions known as \TwoHDMs include a second
complex doublet and are classified into four types according to the
couplings of the Higgs doublets to fermions. The two-doublet structure
gives rise to five physical Higgs bosons via spontaneous symmetry
breaking: two neutral $CP$-even particles \Ph and \PH with $m_{\Ph}\leq m_{\PH}$,
one neutral $CP$-odd particle \PSA, and two charged Higgs
bosons \Hpm. In these models the ratio of the vacuum expectation
values of the two Higgs doublets, \tanbeta, and the mixing angle, \mixingAngle,
between \Ph and \PH  are free parameters. These can be tuned to the
alignment limit \alignmentLimit whereby \Ph aligns with the properties
of the \SM\ Higgs boson with \mh = 125\GeV~\cite{CMS:2020xrn}.

\begin{figure}[!ht]
  \centering
  \includegraphics[width=0.33\textwidth]{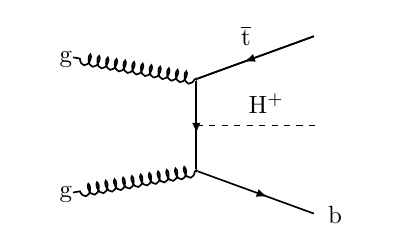}
  \includegraphics[width=0.33\textwidth]{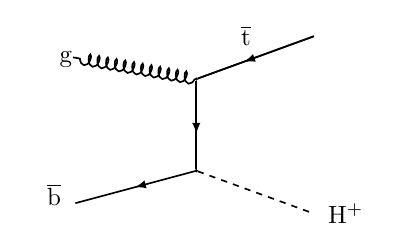}
  \caption{\label{fig:feynman-production}
    Leading order Feynman diagrams for the production of a heavy $\PH^{+}$
    at the LHC through \ppToHplustb in the 4FS (\cmsLeft) and 5FS (\cmsRight).
  }
\end{figure}

Three mass regions are conveniently defined for the classification of
charged Higgs bosons: light (\mHpmLight), intermediate
(\mHpmIntermediate), and heavy (\mHpmHeavy), where \mHpm, \mtop, and \mbottom
represent the masses of the charged Higgs boson, the top and
bottom quarks, respectively. The search described in this paper is focused on a heavy
\PHpm, whose production at the LHC would take place predominantly in association
with top and bottom quarks. The associated top quark production
dominates and can be described in the four- and five-flavor
scheme (4FS and 5FS), which yield consistent results at sufficiently high
order of perturbation theory~\cite{Dittmaier:2003ej,Dawson:2003kb,Harlander:2003ai,Harlander:2011aa}.
The corresponding \LO Feynman diagrams are shown in
\reffig{feynman-production}, with charge-conjugate processes implied.

When considering \TwoHDMs, the decay \BRs of \PHpm can vary significantly
between different models.
Under the enforcement of the Z2 symmetry there are four types of
\TwoHDMs which, for $\tanbeta=1$, lead to the channels \HpmToTauNu and
\HpmToCS being dominant in the light-\PHpm region.
In the heavy-$\Hpm$ region, the decay mode
\HpmToTB dominates, with some competition from the \HpmTohW  and
\HpmToHW decay modes. This behavior also holds in the alignment limit
with the only difference being that \BRto{\HpmToHW} increases faster
with \mHpm, while \BRto{\HpmTohW} vanishes completely.
At larger values of \tanbeta and for \nonAlignmentLimit, the
interplay between the channels \HpmToHW and \HpmTohW becomes more
intricate, with the former becoming important once kinematically
attainable and the latter dominating because of the large phase space available.

The importance of the potential interference between the
\HpmToHW and \HpmTohW channels is difficult to
quantify as it highly depends on the considered parameter space.
In general, however, a large \HpmTohW coupling is always associated with a
small \HpmToHW coupling, and vice versa. As discussed in
\refcite{Bahl:2021str}, assuming a misalignment of $\approx$0.1
with $\sinBetaMinusAlpha=0.9$, \BRto{\HpmTohW} is suppressed by a
factor of $\approx$100 with respect to \BRto{\HpmToHW}, while the
interference between \HpmToHW and \HpmTohW is also expected to be suppressed
by at least a factor of $\approx$10 compared to the contribution from \HpmToHW itself.
In this paper, we focus on the \HpmToHW
decay mode and neglect completely the \HpmTohW decay mode, as well as
their interference. For the neutral Higgs boson, typically the most frequent
final states close to the alignment limit are the $\PQb\PQb$ and $\PGt\PGt$, while the
$\PW\PW$, $\PZ\PZ$, and $\cPgg\cPgg$ channels are experimentally the
cleanest ones. However, the aforementioned decay rates are
model-dependent and are also affected, directly or indirectly, by the
value of \mH.

Since no charged scalar boson exists in the \SM, a
discovery of a charged Higgs boson would provide
unequivocal proof of physics beyond the \SM. To date, various searches for an \PHpm
signature have been conducted by the ATLAS and CMS Collaborations in
\pp collisions at $\sqrt{s}=7$, 8, and 13\TeV.
Searches for a light \PHpm include the channels
\HpmToTauNu~\cite{Aad:2014kga,Khachatryan:2015qxa,Aaboud:2018gjj,Sirunyan:2019hkq},
\HpmToCS~\cite{Khachatryan:2015uua,Aad:2013hla},
\HpmToCB~\cite{Sirunyan:2018dvm}, and
\HpmToWA~\cite{Sirunyan:2019zdq}. In the heavy-\Hpm region, the
searches include the channels \HpmToTB~\cite{Khachatryan:2015qxa,Aad:2015typ,Aaboud:2018cwk,Sirunyan:2019arl}
and \HpmToTauNu~\cite{Aad:2014kga,Khachatryan:2015qxa,Aaboud:2018gjj,Aaboud:2016dig,Sirunyan:2020hwv}.
Charged-current processes from low-energy flavor observables, such as
tauonic \PB meson decays and the \BToSGamma transition, have yielded indirect lower
limits on \mHpm~\cite{Bevan:2014iga,Kou:2018nap}. Searches for \HpmToWZ
decays predicted in Higgs triplet models~\cite{Senjanovic:1975rk,Gunion:1989ci,Georgi:1985nv}
have also been conducted in the vector boson (\VjetsDef) fusion
production mode~\cite{Aad:2015nfa,Sirunyan:2017sbn,Sirunyan:2017ret,CMS:2021wlt}.
Finally, the ATLAS Collaboration has set limits on \PHpm production
with a search for dijet resonances in events with an isolated charged
lepton~\cite{Aad:2020kep}. No evidence of a charged Higgs boson has
been reported in any of the aforementioned searches.
Searches for additional heavy neutral Higgs bosons have also been
performed at experiments at LEP~\cite{Schael:2006cr} and
the Tevatron~\cite{Aaltonen:2009vf,Abazov:2010ci,Abazov:2011jh,Aaltonen:2011nh}.
These are superseded by searches performed by the ATLAS and CMS
Collaborations  in the $\PQb\PQb$~\cite{Chatrchyan:2013qga,Khachatryan:2015tra,Sirunyan:2018taj,ATLAS:2019tpq}, $\PGm\PGm$~\cite{Aad:2012cfr,CMS:2015ooa,CMS:2019mij,ATLAS:2019odt}, and $\PGt\PGt$~\cite{Aad:2012cfr,Aad:2014vgg,Aaboud:2016cre,Aaboud:2017sjh,Chatrchyan:2011nx,Chatrchyan:2012vp,Khachatryan:2014wca,Sirunyan:2018zut,ATLAS:2020zms} final states.

In this paper, a direct search for a heavy \PHpm is performed through the
\HpmToHW and \HToTT decay modes, targeting the \HpmToHW decay channel for the first time at the LHC.
In this search it is assumed that the \PH boson has a mass of $\mH=200\GeV$.
This particular choice appears as a benchmark point in various
scenarios such as extended Inert Doublet Models aiming to
provide a viable dark matter
candidate~\cite{Cordero:2017owj,Cordero-Cid:2018man}, in
\TwoHDM frameworks with new sources of CP-violation~\cite{Keus:2015hva}
or a strong first order electroweak phase
transition~\cite{Kainulainen:2019kyp} which are needed for a
successful electroweak baryogenesis. Such heavy neutral scalars have
also been proposed to address flavor puzzles such as the anomalous
magnetic moment of the muon~\cite{Keus:2017ioh}.

\begin{figure}[!ht]
  \centering
  \includegraphics[width=0.45\textwidth]{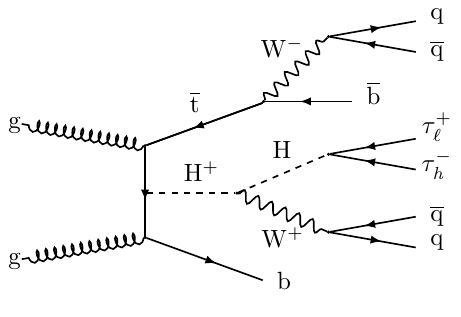}
  \includegraphics[width=0.45\textwidth]{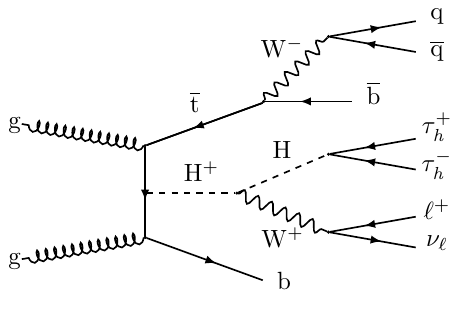}
  \caption{\label{fig:feynman-finalStates}
    Feynman diagrams showing the signal processes targeted by this
    analysis, with the production of a heavy $\PH^{+}$ in the
    4FS, followed by the \HpToHW and \HToTT decays, resulting in
    \ltau (\cmsLeft) and \ltautau (\cmsRight) final states.
    Contributions to the \ltau final state may also arise from the
    \cmsRight diagram when one \tauh from the \HToTT decay is not
    reconstructed.}
\end{figure}

The search focuses on an associated \PHpm production with a
hadronically decaying top quark, in final states with at least one \taujet
and exactly one isolated \lepton, as shown in \reffig{feynman-finalStates}.
Four distinct final states are targeted: \myFS.
For the \ltau final states, candidate events contain one
\tauh candidate, one isolated lepton, missing transverse momentum
(\ptvecmiss), and three additional hadronic jets from \PW boson decays
and \PQb quarks. The \ltau search employs a \MVA classifier based on \BDTG to
distinguish the signal from backgrounds. For the \ltautau final
states, candidate events are selected by
requiring one additional \tauh candidate and by relaxing the hadronic jet multiplicity
requirement to at least two. In these final states, the transverse mass
of the charged Higgs boson, \mT, is used
to distinguish signal from backgrounds. Upper limits on the product of
the \PHpm production cross section, $\sigma_{\PHpm}$, and the
branching fraction $\BRto{\HpmToHW, \HToTT}$ for the decay chain
\HpmToHW with \HToTT, are presented as functions of \mHpm.

The paper is organized as follows. A brief description of the CMS
detector is given in \refsec{detector}, while the collision data and
simulated samples are discussed in \refsec{samples}. \refsec{reconstruction}
describes the global event reconstruction and physics object
identification, followed by the  event selection in \refsec{selections}.
Background estimation, search strategy, and systematic uncertainties
are described in \reftrisec{background}{strategy}{systematics}, respectively.
Finally, the results are presented in \refsec{results} and summarized
in \refsec{summary}.

\section{The CMS detector}
\label{sec:detector}

The central feature of the CMS apparatus is a superconducting solenoid
of 6\unit{m} internal diameter, providing a magnetic field of
3.8\unit{T}. Within the solenoid volume are a silicon pixel and strip
tracker, a lead tungstate crystal \ECAL,
and a brass and scintillator \HCAL, each composed of a
barrel and two endcap sections. Forward calorimeters extend the
pseudorapidity ($\eta$) coverage provided by the barrel and endcap
detectors up to $\abs{\eta} = 5$. Muons are detected in gas-ionization
chambers embedded in the steel flux-return yoke outside the solenoid.

Events of interest are selected using a two-tiered trigger
system. The first level, composed of custom
hardware processors, uses information from the calorimeters and muon
detectors to select events at a rate of around 100\unit{kHz} within a
time interval of less than 4\mus~\cite{Sirunyan:2020zal}.
The second level, known as the \HLT, consists of a farm of processors running a
version of the full event reconstruction software optimized for fast
processing, and reduces the event rate to around 1\unit{kHz} before
data storage~\cite{Khachatryan:2016bia}.

A more detailed description of the CMS detector, together with a
definition of the coordinate system used and the relevant kinematic
variables, can be found in \refcite{Chatrchyan:2008zzk}.

\section{Collision data and simulated samples}
\label{sec:samples}

The analysis presented in this paper uses \pp collision data collected with the
CMS experiment at $\sqrt{s}=13\TeV$ during the years 2016, 2017, and 2018. The
corresponding integrated luminosities are 36.3, 41.5, and
59.8\fbinv, respectively, amounting to a total of \intLumi. The
aforementioned data were collected with the use of single-electron and
single-muon triggers with isolation criteria. The trigger thresholds
are mentioned in \refsec{selections}.

\sloppy
Simulated events are used to model the signal and background processes
using various \MC event generators. The signal samples are generated with the
\MGvATNLO generator~\cite{Alwall:2014hca} v2.2.2 for 2016 (v2.4.2 for
2017--2018) using the 4FS at \NLO precision in \QCD.
The decays of the \PHpm resonances are generated with
\MADSPIN to preserve spin-correlation and finite-width
effects. Both charge-conjugate signal processes are generated with
four mass hypotheses $\mHpm = 300$, 400, 500, and 700\GeV and
under the assumption that $\mh = 125\GeV$ and $\mH = 200\GeV$.

The \SMttbar ($\ttbar$) production constitutes an important background that contributes
significantly to all final states considered. It is simulated with
\NLO precision in \QCD using the \POWHEGtwo generator. Its cross
section is obtained from the \textsc{TOP++}~v2.0~\cite{Czakon:2011xx}
calculation that includes \NNLO corrections in \QCD and resummation
of the next-to-next-to-leading logarithmic soft-gluon terms.  Other important sources of background
include single top quark production (\SingleTop), \ttX with
$\PX=\PW$, \PZ, \Ph, or \ttbar, and \EW. For the \SingleTop samples, the $t$-channel
process is generated with \POWHEGtwo at \NLO precision in \QCD using
the 4FS~\cite{Frederix:2012dh} and interfaced with \MADSPIN for
simulating the top quark decay. The $s$-channel process is simulated
using \MGvATNLO, while the production via the $\PQt\PW$-channel is
simulated at \NLO in \QCD using the 5FS and
\POWHEGtwo~\cite{Re:2010bp}. The production of \ttbar in association
with \PW or \PZ boson is simulated with \NLO\ precision in
\QCD using \MGvATNLO. The $\ttbar\Ph$ background process is generated
using \POWHEGtwo at \NLO~\cite{Hartanto:2015uka}, with a Higgs boson
mass of 125 \GeV. The top quark mass is set to $\mtop=172.5\GeV$.
The \Vjets samples  are generated at \LO precision using \MGvATNLO, with up to four
partons included in the \ME calculations. Finally, the
\Diboson processes are generated at \LO precision using \PYTHIAeight.

For processes generated at \LO precision in \QCD, the \MLM matching
and merging procedure is used~\cite{Alwall:2007fs}, whereby partons
from the \ME calculation are matched to the jets reconstructed after
the perturbative shower. For processes generated at \NLO precision in
\QCD with the \MGvATNLO generator, the events from the
\ME characterized by different parton multiplicities are merged with
the FxFx procedure~\cite{Frederix:2012ps}. The matching between the
\ME calculation and the parton shower in \POWHEGtwo is controlled by
the damping factor $\hdamp$, which has a value set to $\hdamp=1.379 \mtop$.
It is used to limit the resummation of higher-order effects by the
Sudakov form factor to below a given \pt scale.

For the generation of the
above-mentioned simulated processes, the \PDFs are parameterized
using NNPDF3.0~\cite{Ball:2014uwa} for 2016
(NNPDF3.1~\cite{Ball:2017nwa} for 2017--2018). The \PDFs in the
\ME calculations are at \NLO for  NNPDF3.0 and at \NNLO for NNPDF3.1. The
parton shower and fragmentation are modeled with the
\PYTHIA generator v8.212 for 2016 samples (v8.230 for 2017--2018).
The \PYTHIA parameters affecting the description of the
\UE are set to the {CUETP8M1} tune~\cite{Khachatryan:2015pea} for
2016 ({CP5} tune~\cite{TuneCP5} for 2017--2018). The
response of the CMS detector is simulated using
\GEANTfour~v9.4~\cite{Agostinelli:2002hh} and reconstructed using the
same version of the CMS software as that used for
the collision data.

The effect of additional inelastic \pp interactions
within the same or nearby bunch crossings, henceforth referred to as
pileup, is taken into account by generating concurrent minimum bias events.
All simulated events are weighted to match the pileup distributions
observed in the data. The average number of pileup in the 2016 data
set was 23, increasing to 32 during the 2017--2018 data taking.

\section{Object reconstruction}
\label{sec:reconstruction}

The global event reconstruction is based on the \PF
algorithm~\cite{CMS:2017yfk}, which uses an optimized combination
of information from the elements of the CMS detector to reconstruct
individual particles in an event. It categorizes these \PF candidates
as photons, electrons, muons, charged hadrons, and neutral hadrons.
Higher-level objects are reconstructed from combinations of the
\PF\ candidates.
The \PVtx\ is taken to be the vertex corresponding to the hardest
scattering in the event, evaluated using tracking information alone.
More specifically, the individual tracks originating from the same
candidate vertex are clustered using the anti-\kt algorithm with a
distance parameter of 0.4~\cite{Cacciari:2008gp}, as implemented in
the \FASTJET library~\cite{Cacciari:2011ma}. For each \PVtx candidate
the $\sum\pt^{2}$\ value is computed by considering the clustered
jets, the remaining single tracks, and the neutral particle
contributions inferred from the negative vector sum of the \pt of
those jets. The \PVtx with the largest $\sum\pt^{2}$ is chosen as the
one corresponding to the hard scattering. All other candidate
vertices are attributed to pileup, with the exception of secondary vertices that
are transversely displaced from the \PVtx and indicative of
decays of long-lived particles emerging from it.

Electrons are identified as charged-particle tracks that are
potentially associated with \ECAL energy clusters and bremsstrahlung
photons emitted while traversing the tracker material.
Their momentum is estimated by combining the energy measurement in the
\ECAL with the momentum measurement in the tracker. The momentum
resolution for electrons with $\pt \approx 45\GeV$ from $\Z
  \rightarrow \Pe \Pe$ decays ranges from 1.7 to 4.5\%.
It is generally better in the barrel region than in the endcaps, and
also it depends on the bremsstrahlung energy emitted by the electron as
it traverses the material in front of the \ECAL.
An \MVA discriminant~\cite{CMS:2020uim} is used to achieve better discrimination of prompt
isolated electrons from other electron candidates, mainly originating from
photon conversions, jet misidentification, and semileptonic \PQb hadron decays.
It requires as input several
variables describing the shapes of the energy deposits in the
\ECAL and the track quality. In this paper, a medium (loose) working
point with an identification efficiency of 90 ($>$99)\% is used for selecting
(vetoing) electrons, corresponding to a rate of jets misidentified as
electrons of $\approx$1 (20)\%.

Muons are reconstructed as tracks in the central tracker consistent with
either a track or several hits in the muon system, and associated with
calorimeter deposits compatible with the muon hypothesis.
They are measured in the range of $\abs{\eta} < 2.4$,
with detection planes using three technologies: drift tubes,
cathode strip chambers, and resistive plate chambers.
Their momentum is obtained from the curvature of the corresponding
tracks in the silicon tracker, with a relative resolution of 1 and 3\%
for muons with \pt up to 100\GeV in the barrel and endcaps,
respectively. The \pt resolution in the barrel is better than 7\% for
muons with \pt up to 1\TeV~\cite{CMS:2018rym}.
To increase the purity of prompt muons originating at
the \PVtx, a set of discriminants is employed based on the track
fit quality, the number of hits per track, and the degree of
compatibility of the information from the tracker and muon systems.
A tight (loose) working point with an efficiency of $\approx$95 (99)\%
is used for selecting (vetoing) muons, in order to suppress muons from
decays in flight and misidentified muons from hadronic punch-through.

Both prompt and displaced reconstructed leptons are used in this
analysis, with the latter exclusively used in the validation of
the background estimation, described in detail in \refsec{background}.
In both cases the background contributions from misidentified
\leptons are further suppressed by applying stringent requirements on
the lepton isolation. The relative lepton isolation variable
\rIso{} is employed to ensure that the leptons are not associated with
any significant electromagnetic or hadronic activity in the
detector. It is defined as the scalar \pt sum, normalized to the
lepton \pt, of all \PF candidates within a cone of radius $\Delta R =
  \sqrt{\smash[b]{(\Delta\eta)^2 + (\Delta\phi)^2}} < 0.3$ (0.4)
around the electron (muon) direction at the \PVtx,
where $\phi$ is the azimuthal angle in radians.
The lepton itself is excluded from the calculation.
To mitigate any effects due to
contamination from pileup, only \PF candidates whose tracks are
associated with the \PVtx are taken into account. For neutral hadrons
and photons, where the absence of an associated track precludes an
unambiguous association with the \PVtx, an estimate of the pileup
contribution is subtracted from their energy
sums~\cite{CMS:2020ebo}. A tight (loose) isolation criterion with
discriminant $\rIso{}<0.15$ (0.25) is used in lepton selection
(veto).  The three-dimensional impact parameter significance, which is
the impact parameter value normalized to its uncertainty, can also be
used to further suppress electrons from photon conversions and muons
originating from in-flight decays of hadrons. Its value is required to
be less than ten for prompt electrons, or greater than three for all
displaced electrons or muons.

Jets are reconstructed from \PF candidates using the anti-\kt
algorithm~\cite{Cacciari:2008gp}, as implemented in the \FASTJET
package~\cite{Cacciari:2011ma}, with a distance parameter of 0.4.
The jet momentum is determined as the vectorial
sum of all particle momenta in the jet. It is found from simulation
to be within 5--10\% of the true momentum, over the whole \pt
spectrum and detector acceptance. Pileup can contribute additional
tracks and calorimetric energy depositions, increasing the apparent
jet momentum. To mitigate this effect, tracks identified to be
originating from pileup vertices are discarded and an offset
correction is applied to correct for remaining
contributions~\cite{CMS:2020ebo}. Jet energy corrections are derived
from simulation studies so that the average measured energy of jets
becomes identical to that of particle-level jets. In situ measurements
of the momentum balance in dijet, photon+jet, \PZ{+}jet, and multijet
events are used to determine any residual differences between the jet
energy scale in data and in simulation, and appropriate corrections
are applied to simulated events~\cite{CMS:2016lmd}. After the corrections,
the jet energy resolution amounts
typically to 15--20\% at 30\GeV, 10\% at 100\GeV, and 5\% at
1\TeV~\cite{CMS:2016lmd}.

The identification of jets that originate from the hadronization of \PQb
quarks (\bjets) is performed with the \DEEPJET multiclass
flavor-tagging algorithm, as described in
\refcites{Bols:2020bkb,CMS-DP-2018-058, BTV-16-002,Bols:2020bkb}.
In this analysis, a medium working point of this algorithm is
chosen that corresponds to a \bjetAdj identification efficiency of
$\approx$80\%. The associated misidentification rate for jets
originating from light quarks and gluons (\PQc quarks) is 1
(15)\%~\cite{CMS-DP-2018-058}.

The \tauh candidates are reconstructed with the
hadrons-plus-strips algorithm, as described in \refcite{Sirunyan:2018pgf}.
It uses clustered anti-\kt jets as seeds to reconstruct \PGt decay
modes with one charged hadron and up to two neutral pions (one-prong),
or three charged hadrons and up to one neutral pion (three-prong). The
neutral pions, which decay promptly to a photon pair, are
reconstructed as strips of \pt-dependent size in the \etaphi plane from
reconstructed electrons and photons contained in the jet.
These strips are narrow in $\eta$ but wide in $\phi$
to allow for the broadening of \ECAL energy deposits due
to photon conversions. The \taujet decay mode is then obtained
by combining the charged hadrons with the strips, resulting in a reconstruction
efficiency of $\approx$80\%. To efficiently discriminate the \taujet decays
against jets originating from the hadronization of quarks or gluons, and
against electrons or muons, the \DEEPTAU multiclass \PGt identification
algorithm is used. It exploits the reconstructed event quantities by combining
low-level information from the tracking, calorimeter, and muon
detectors with high-level properties of the \taujet\ candidate and other
\PF candidates in its vicinity. The multiclassification
output \yDeepTauDef represents a Bayesian probability that the \taujet candidate
originates from a \lepton ($\alpha=\Pe$, \PGm), the hadronization
of a quark or gluon ($\alpha=j$), or a genuine \taujet
($\alpha=\PGt$). The aforementioned output enables the definition of
three discriminators according to the ratio
$\dDeepTau{\alpha} = \yDeepTau{\PGt}/(\yDeepTau{\PGt}+\yDeepTau{\alpha})$
with $\alpha=\Pe$, \PGm, $j$.
For this analysis, medium and tight working points of $\dDeepTau{\Pe}$
and $\dDeepTau{\PGm}$ are used with efficiencies of 62 and 70\% and
misidentification rates of 0.2 and 0.03\%, respectively.
For \dDeepTau{j}, the medium and loose
working points are used with efficiencies of
49 and 70\% for misidentification rates of 0.4 and 5\%,
respectively, for \taujet candidates with \pt\ up to 200\GeV. The
selected \tauh candidates that pass
the loose but fail the medium \dDeepTau{j} working point are referred
to as anti-isolated \tauh's and are solely used in the background
estimation as described in \refsec{background}.

The \ptvecmiss is computed as the negative vector sum of the
transverse momenta of all \PF candidates in an event, and its
magnitude is denoted as \ptmiss~\cite{Sirunyan:2019kia}. The
\ptvecmiss is modified to account for corrections to the energy
scale of the reconstructed jets in the event. The scalar \pt sum of
all selected leptons and \tauh objects in an event is denoted as \LT,
while the corresponding sum over all selected jets is designated as
\HT. In addition, we define \ST as the scalar sum of \ptmiss, \LT, and
\HT in the event. The transverse mass of charged Higgs boson
candidates is calculated as:
\begin{linenomath}
  \begin{equation}
    \label{eq:mt}
    \mT = \sqrt{ (\ET^{1} + \ET^{2} + \ET^{\PW} + \ptmiss)^{2}
      - \abs{\ptvec^{1} + \ptvec^{2} + \ptvec^{\PW} + \ptvecmiss}^{2}},
  \end{equation}
\end{linenomath}
where $\ET^{1}$, $\ET^{2}$, and $\ET^{\PW}$ are the total visible
transverse energies of the two tau lepton and {\PW} boson decay products,
respectively, and $\ptvec^{1}$, $\ptvec^{2}$, and $\ptvec^{\PW}$ the
corresponding transverse momentum vectors.

Hadronic decays of top quarks are reconstructed and identified using
a \resTop tagger that is based on a fully connected neural network
implemented using \textsc{Keras}~\cite{chollet2015keras} and
\textsc{TensorFlow}~\cite{tensorflow} software packages. It targets
top quarks whose decay products are resolved as three separate
anti-\kt jets with a distance parameter of 0.4. It is
trained on simulated \ttbar events to discriminate between three-jet
combinations originating from the decay of top quarks (signal)
and other combinatorial three-jet systems (background).
For training, the signal
\resTop candidates are matched to the generated top quark decay
products with one-to-one jet-to-quark matching, while the
\resTop candidates with at least one unmatched jet are considered
as combinatorial background.

This \MVA classifier utilizes high-level information from each of the
three seed jets, such as invariant masses, angular separations, jet
flavor, and jet shape variables. All the selected variables are
uncorrelated with the top quark mass to minimize possible correlations
between the \resTop candidate mass and the associated classifier
output. To improve the stability and performance of the learning
algorithm, the input features are transformed such that they are
distributed in similar ranges and not influenced by outliers. This
tagger uses the robust scaler preprocessing method via
\textsc{scikit-learn} tool~\cite{scikit-learn}. Furthermore to prevent
mass sculpting effects, the algorithm uses the sample-reweighting
technique to decorrelate the classifier's output from the top quark
mass. The mass information is removed from all the input features by
reweighting the combinatorial background data sets so that the
\resTop candidate mass distribution matches that of the signal.

\begin{figure}[bth]
  \centering
  \includegraphics[width=0.7\linewidth]{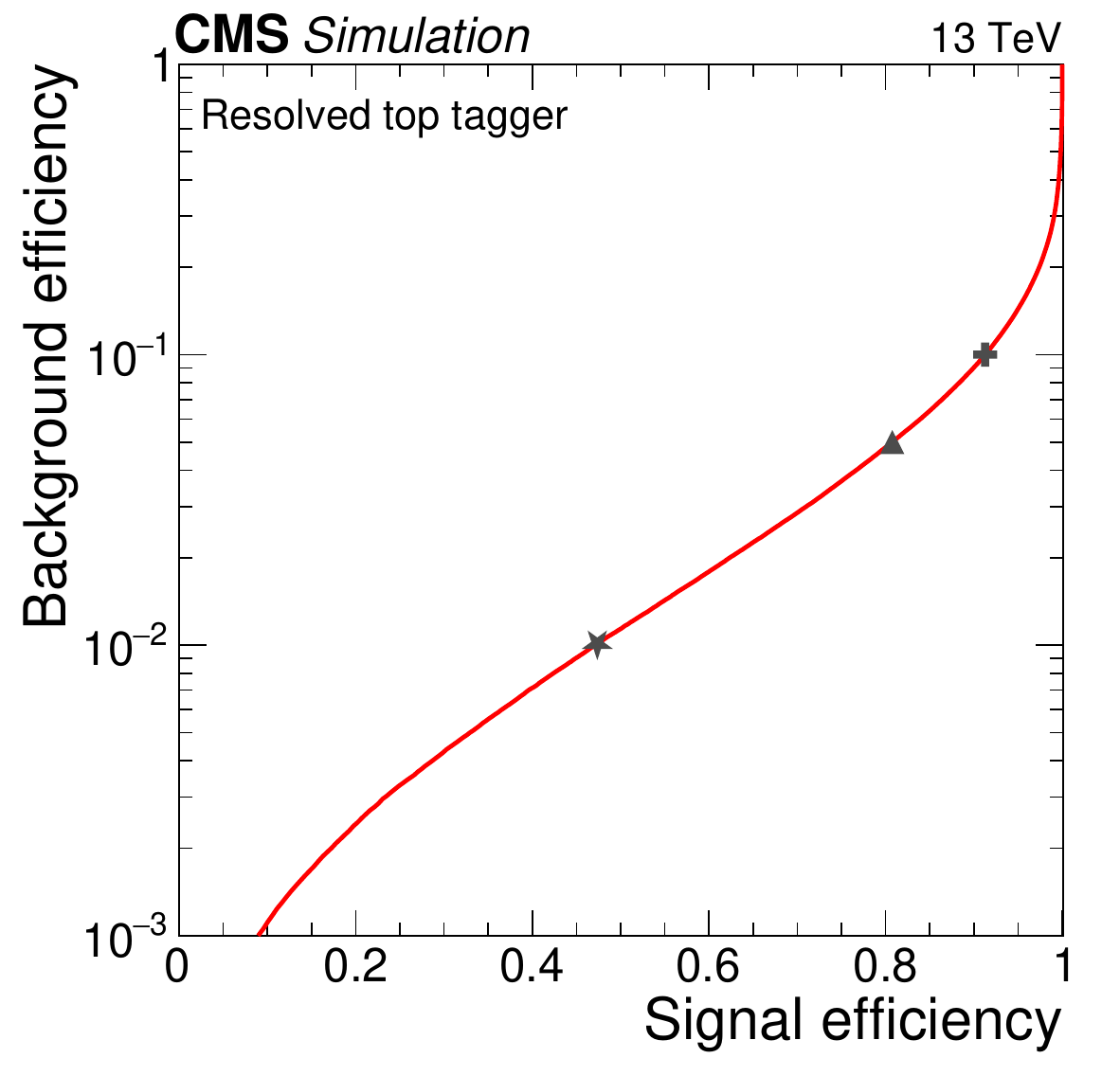}
  \caption{Receiver operating characteristic curve of the
    \resTop tagger. The cross-, triangle-, and star-shaped markers
    indicate the loose, medium, and tight working points  with 10,
    5, and 1\% background misidentification probability. The
    corresponding identification efficiencies are 91, 81, and
    47\%, respectively.}
  \label{fig:topTag-ROC}
\end{figure}

The performance of the \resTop tagger is expressed as a receiver
operating characteristic curve and it is shown in
\reffig{topTag-ROC}. The loose, medium, and tight working points are
established at 10, 5, and 1\% background misidentification
probability. The corresponding identification efficiencies are
91, 81, and 47\%, respectively. The loose working point
is employed in this analysis.

\begin{figure}[ht!]
  \centering
  \includegraphics[width=.49\linewidth]{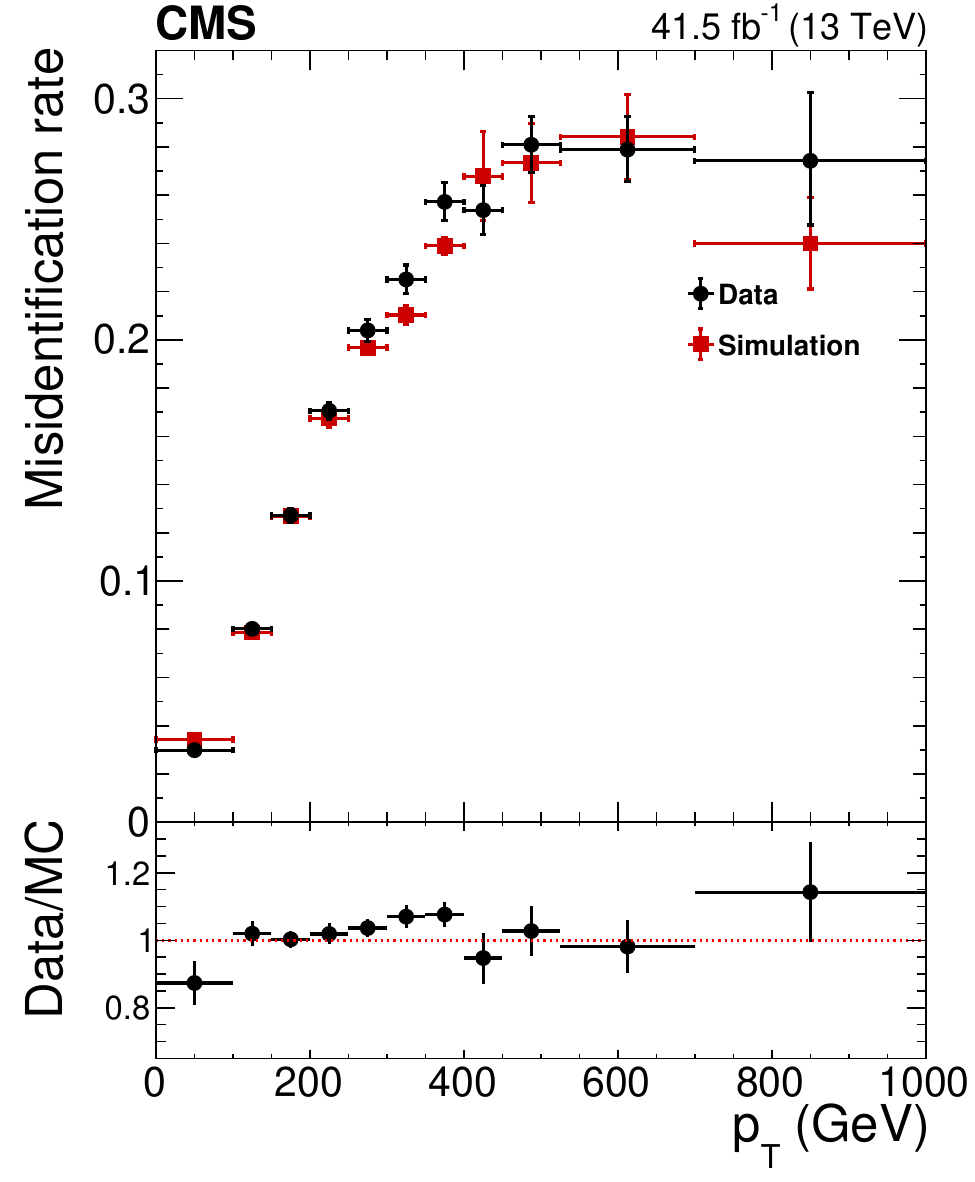}
  \includegraphics[width=.49\linewidth]{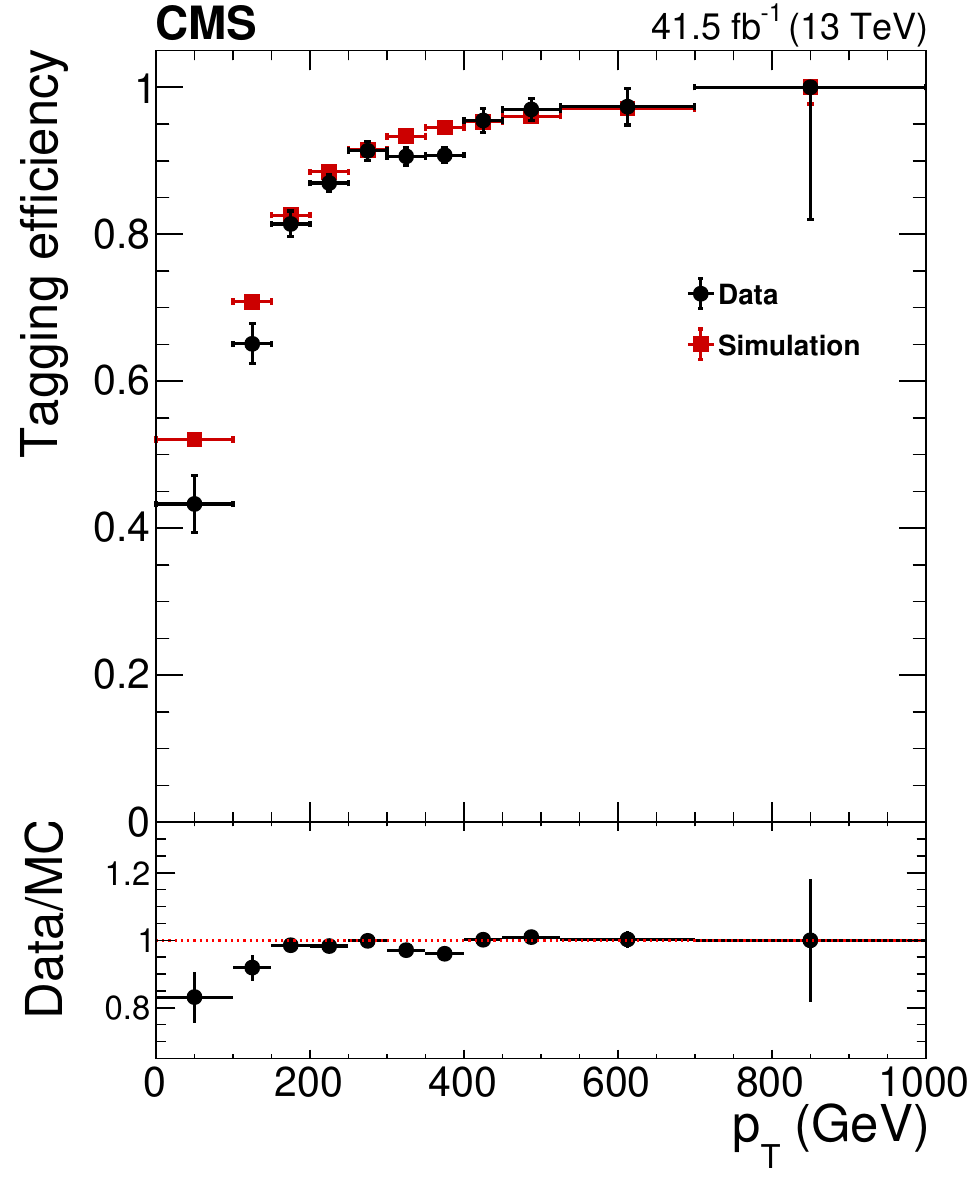}
  \caption{
    Misidentification rate (\cmsLeft) and \resTop-tagging efficiency (\cmsRight)
    in data and simulation,  as a function of the \resTop candidate
    \pt for the loose working point, using the
    2017 data.
  }
  \label{fig:topTag-mistagSF}
\end{figure}

The misidentification rate and tagging efficiency of the \resTop tagger
have been estimated and compared in data and simulation to
extract data-to-simulation corrections, using a sample of
lepton+jets, dominated by the semileptonic \ttbar events. The events
are characterized by large \ptmiss, exactly one muon
identified as tight with $\pt>50\GeV$ and at least four jets with
$\pt>40\GeV$ and $\abs{\eta}<2.4$, of which at least one is
b tagged. The jet closest to the muon is considered as the b jet from
the leptonic top quark decay.
The three-jet system with a mass closest to the top quark mass
is selected as the hadronic \resTop candidate.
The misidentification rate is measured using events in which the
\resTop candidate mass, \mresTop, falls outside a mass window of
130--210\GeV, and the sample is dominated by the combinatorial
background. Events where \mresTop is within the mass window are used
to measure the \resTop tagging efficiency, after subtracting from data
the contributions from non-top quark processes and the combinatorial
background, as estimated from simulations.
The misidentification rate and tagging efficiency
for the 2017 data set are shown in \reffig{topTag-mistagSF} as a
function of the \resTop candidate \pt for the loose working point.
The data-to-simulation corrections are defined as the ratio
data/simulation, in \pt bins of the \resTop candidate.
Similar behavior was also observed for the 2016 and 2018 data.

\section{Event selection}
\label{sec:selections}
The analysis is conducted in the mutually exclusive \myFS final states.
The event selection strategy is independently optimized for each final
state to improve the suppression of backgrounds while maintaining a
large signal selection efficiency. The selection of signal candidate
events at the trigger level is based on the presence of at least a
single isolated lepton. More specifically, for the \etau and
\etautau channels the online \HLT requires the presence of an
isolated electron with a \pt threshold of 27, 32, and 32\GeV for the
data-taking years 2016, 2017, and 2018, respectively. For the
\mtau and \mtautau channels, an isolated muon  is required,
with \pt thresholds of 24, 27, and 24\GeV for the three years.
The \HLT\ objects are geometrically matched to analogous offline objects that
satisfy the $\pt$, $\eta$, $\rIso{}$ criteria described in the
following, in addition to the object identification requirements given
in \refsec{reconstruction}.

To ensure high trigger efficiency,
the offline $\pt$ and $\eta$ requirements for
prompt electrons (muons) are $\pt > 30$, 35, 35 (26, 29, 26)\GeV
and $\abseta \le 2.1$ (2.4) for the 2016,
2017, and 2018 data-taking periods. Selected electrons (muons) are
also required to pass medium (tight) identification criteria and satisfy
$\rIso{} < 0.15$ for all three years. Only events with exactly one
such electron (muon) are accepted. Furthermore, events with any
additional electrons or muons fulfilling looser
identification criteria are also discarded, provided that
the electron (muon) candidates satisfy
the requirements of $\pt > 10 \GeV$, $\abseta \le 2.4$
(2.5), and $\rIso{} < 0.25$.
This helps to avoid the incorrect assignment of objects in the
transverse mass reconstruction and the inadvertent smearing of its
Jacobian peak, and ensures the
orthogonality between final states with different lepton flavors.

\sloppy
In both the \ltau and \ltautau final states, all \tauh candidates are
required to have $\pt > 20\GeV$ and $\abseta < 2.3$.
They must be well separated in \etaphi space from the trigger
lepton and other \tauh candidates, such that $\dR{\tauh}{\ell/\tauh}>0.5$.
To reduce the contribution of electrons, muons, or jets mimicking a
\tauh\ object, the medium, tight, and medium working points of the \dDeepTau{\Pe},
\dDeepTau{\PGm}, and \dDeepTau{j} discriminants are chosen,
irrespective of final state, or year
of data taking. In the \ltau final states, the events are
classified according to the sum of the electric charge
of the selected lepton and \tauh objects, \charge{\ltauh}, in units of the electron
charge $e$. Both the \oSign case with $\abs{\charge{\ltauh}}=0$ and the \sSign case
with $\abs{\charge{\ltauh}}=2$ are considered as separate
categories.
The categorisation of the \ltauh final states into \oSign and \sSign significantly
improves the expected sensitivity over the entire \mHpm region
considered. Events in the \ltauh \sSign category arise when one of the tau
leptons from the $\HToTT$ decay is not reconstructed or identified. It
has significantly smaller contributions from \SM processes with a top
quark in the final state than the \ltau \oSign category.
In the \ltautau final states, the two \tauh candidates are required to be
\oSign with $\abs{\charge{\tauhtauh}}=0$.

The \ltau final states are required
to contain at least three jets with $\pt>30\GeV$ and
$\abs{\eta} <2.4$, for all data-taking years. Slightly modified
criteria are used for the \ltautau final states, with the multiplicity
requirement reduced to two and the pseudorapidity extended to
$\abs{\eta} < 4.7$, in order to increase the signal acceptance.
All selected jets must be well separated from the reconstructed \tauh
objects such that $\dR{j}{\tauh}>0.5$.
We denote the selected jet with the highest (second-highest) \pt in
each event as $j_1$ ($j_2$). For all final states considered, at least
one of the selected jets is required to pass the medium working point
of the \DEEPJET \bjet identification algorithm.

To further suppress multijet events with nonprompt leptons and jets
misidentified as \tauh, the \ltau (\ltautau) final states require the
presence of moderate magnitude of missing transverse momentum  of
$\ptmiss \geq 40$ (30)\GeV. For the \ltautau final state, the
requirement $\ST > 400\GeV$ is also used as it considerably improves
the expected sensitivity. A summary of the event selection criteria
for all final states is shown in \reftab{selections-SRs}. They result
in a total of 6 \SRs per year of data taking; \etau \oSign,
\etau \sSign, \mtau \oSign, \mtau \sSign, \etautau,
and \mtautau. They are complemented by the various \CRs and \VRs
described in \refsec{background}, which are used to predict the
dominant backgrounds in the statistical inference of the signal.

\begin{table}[!ht]
  \centering
  \topcaption
  {
    Offline selections applied to the reconstructed
    objects to obtain the \SRs of
    the \ltau and \ltautau final states. The \pt, \ptmiss, and \ST
    variables are reported in units of \GeVns, and \charge{} in units of $e$.
    Selection criteria that depend on the year of data taking are
    presented in parentheses with the order
    corresponding to (2016, 2017, 2018). The symbol $\star$ is used
    to represent an electron (muon) for the \etau (\mtau) final
    states, and a \tauh object in the \etautau and \mtautau final states.
  }
  \label{tab:selections-SRs}
  \cmsTable{
    \begin{tabular}{l l l l l l}
      \multicolumn{1}{c}{Object} & \multicolumn{1}{c}{Selection}     & \multicolumn{4}{c}{Signal Regions}                                                                                             \\
                                 &                                   & \multicolumn{1}{c}{\etau}          & \multicolumn{1}{c}{\mtau}   & \multicolumn{1}{c}{\etautau} & \multicolumn{1}{c}{\mtautau} \\
      \hline
      \multirow{4}{*}{Prompt electrons}
                                 & $N$                               & $=$1                               & $=$0                        & $=$1                         & $=$0                         \\
                                 & $\pt$                             & $>$(30, 35, 35)                  & $>$10                       & $>$(30, 35, 35)            & $>$10                        \\
                                 & $\abseta$                         & $<$2.1                             & $<$2.5                      & $<$2.1                       & $<$2.5                       \\
                                 & $\rIso{}$                         & $<$0.15                            & $<$0.25                     & $<$0.15                      & $<$0.25                      \\
      \\
      \multirow{4}{*}{Prompt muons}
                                 & $N$                               & $=$0                               & $=$1                        & $=$0                         & $=$1                         \\
                                 & $\pt$                             & $>$10                              & $>$(26, 29, 26)           & $>$10                        & $>$(26, 29, 26)            \\
                                 & $\abseta$                         & $<$2.4                             & $<$2.4                      & $<$2.4                       & $<$2.4                       \\
                                 & $\rIso{}$                         & $<$0.25                            & $<$0.15                     & $<$0.25                      & $<$0.15                      \\
      \\
      \multirow{4}{*}{\tauh objects}
                                 & $N$                               & \multicolumn{2}{c}{$=$1}           & \multicolumn{2}{c}{$=$2}                                                                  \\
                                 & $\pt$                             & \multicolumn{2}{c}{$>$20}          & \multicolumn{2}{c}{$>$20}                                                                 \\
                                 & $\abseta$                         & \multicolumn{2}{c}{$<$2.3}         & \multicolumn{2}{c}{$<$2.3}                                                                \\
                                 & $\abs{\charge{\ltauhORtauhtauh}}$ & \multicolumn{2}{c}{$=$0, 2}        & \multicolumn{2}{c}{$=$0}                                                                  \\
      \\
      \multirow{3}{*}{Jets}
                                 & $N$                               & \multicolumn{2}{c}{$\geq$3}        & \multicolumn{2}{c}{$\geq$2}                                                               \\
                                 & $\pt$                             & \multicolumn{2}{c}{$>$30}          & \multicolumn{2}{c}{$>$30}                                                                 \\
                                 & $\abseta$                         & \multicolumn{2}{c}{$<$2.4}         & \multicolumn{2}{c}{$<$4.7}                                                                \\
      \\
      \multirow{3}{*}{\bjets}
                                 & $N$                               & \multicolumn{2}{c}{$\geq$1}        & \multicolumn{2}{c}{$\geq$1}                                                               \\
                                 & $\pt$                             & \multicolumn{2}{c}{$>$30}          & \multicolumn{2}{c}{$>$30}                                                                 \\
                                 & $\abseta$                         & \multicolumn{2}{c}{$<$2.4}         & \multicolumn{2}{c}{$<$2.4}                                                                \\
      \\
      \multirow{2}{*}{Event-based}
                                 & $\ptmiss$                         & \multicolumn{2}{c}{$>$40}          & \multicolumn{2}{c}{$>$30}                                                                 \\
                                 & $\ST$                             & \multicolumn{2}{c}{\NA}            & \multicolumn{2}{c}{$>$400}                                                                \\
      \hline
    \end{tabular}
  }
\end{table}

\section{Background estimation}
\label{sec:background}
The dominant background for all the final states considered stems
from \Vjets and \SMttbar productions. This can be decomposed into events with
\genuineTauh candidates, and events with leptons or jets misidentified as
\tauh candidates. Backgrounds from events with genuine \tauh candidates, together with
events involving electrons or muons misidentified as \tauh objects
(\LToTauh) are estimated from simulation. The reconstructed \tauh candidate is
matched to a generator-level tau lepton, electron, or muon using a cone of
$\Delta R=0.1$. Backgrounds from jets misidentified as \tauh
candidates (\JetToTauh) are estimated using control samples in data by
the use of misidentification rates. These misidentification rates are
measured in dedicated \CRs  that are enriched in jets misidentified as
\tauh candidates, and mimic as closely as possible the composition and
kinematic properties of the corresponding \SRs. The \CRs are required
to be orthogonal to all \SRs and have negligible signal
contamination. Then, the misidentification rate is defined as:

\begin{linenomath}
  \begin{equation}
    \label{eq:fake-rate}
    \tauFR = \frac{N^{\text{CR}}_{\text{nominal } \tauh}}{N^{\text{CR}}_{\text{loose},
          \tauh}}
  \end{equation}
\end{linenomath}
where $N^{\text{CR}}_{\text{nominal } \tauh}$  is the number of events in the \CR\
satisfying the nominal \tauh selection criteria, and
$N^{\text{CR}}_{\text{loose } \tauh}$ is the corresponding number of events in
the \CR satisfying loose \tauh selection criteria.
For both the numerator and denominator in \refeq{fake-rate} a
correction is applied to remove events containing jets originating
from a \genuineTauh candidate or from a lepton misidentified as a \tauh candidate.

The misidentification rate measurements are performed in different
\CRs, separately for each final state and data-taking period. In the
\ltau channels, the \CRs are selected with the nominal criteria used
in defining the \SRs, but with the following modifications.
The \ptmiss criterion is inverted such that $\ptmiss < 40\GeV$, while
the requirement of exactly one \rTop object
is also introduced to suppress contributions from \Vjets
processes. The motivation for these selections is
twofold: to enforce orthogonality with the \SRs, and to ensure that the
obtained regions are enriched in \ttbar. A second set of \CRs, orthogonal
to all other \CRs\ and \SRs, is also used with alternative selection
criteria that are introduced to enhance \EW contributions. More
specifically, the \ptmiss requirement is removed, the \bjet multiplicity is inverted,
and the requirement of exactly one \rTop object is also imposed.
These auxiliary \CRs are used to quantify systematic uncertainties
related to differences in sample composition between the \SRs and the
\CRs, and in particular the relative contribution of \SMttbar and
\EW processes.
The misidentification rates are measured as a function of the \tauh
candidate \pt, separately for one- and three-prong decays, and individually
for the central ($\abseta \le 1.5$) and forward regions
($\abseta\geq1.5$) of the detector.

In the \ltautau channels, the misidentification rates are measured using similar
event selection as in the \SRs, except that the two \tauh objects are
required to be \sSign with $\abs{\charge{\tauhtauh}}=2$, and no
requirement is imposed on the \ST variable.
To compensate for the lower statistical precision relative to the
\ltau final states, the misidentification rates are parametrized in
the \pt and decay mode of the \tauh objects, without separating
the central and forward regions of the detector.

The predicted number of events with \fakeTauh objects in the \SRs is
derived using the fake factor method~\cite{Sirunyan:2018qio} by
applying the misidentification rates evaluated in the \CRs to events
with anti-isolated \tauh objects:
\begin{linenomath}
  \begin{equation}
    \label{eq:fake-rate-prediction}
    N_{\JetToTauh}^{\text{SR}} = \sumBins{i} w_{i}^{\text{CR}} N_{\text{one anti-isolated } \tauh,i}^{\text{SR}}
    - \sumBins{i} w_{i,~1}^{\text{CR}} w_{i,~2}^{\text{CR}} N_{\text{two anti-isolated} \tauh,i}^{\text{SR}}.
  \end{equation}
\end{linenomath}
The index $i$ refers to each bin of the parametrization and $w_{i}$ to
the corresponding normalization weight. The term $N_{\text{anti-isolated } \tauh,i}^{\text{SR}}$ refers to
number of events with anti-isolated \tauh objects in the \SR of interest,
after subtracting events with jets originating from a
\genuineTauh\ or from a lepton misidentified as a \tauh candidate.
The normalization weight for a given parametrization bin $i$ is given by:
\begin{linenomath}
  \begin{equation}
    \label{eq:fake-rate-weight}
    w_{i}^{\text{CR}}  = \frac{\tauFRindex{i}}{1-\tauFRindex{i}}
  \end{equation}
\end{linenomath}
for each event with a single anti-isolated \tauh candidate present.
In the \ltautau final states, \refeq{fake-rate-weight} must be
applied to each of the two \tauh candidates that are present. It thus
accounts for the case where only one of the \tauh candidate is a
\fakeTauh, but also includes cases whereby both \tauh candidates are
misidentified and one of them passes all nominal identification criteria.
In order to account for this double counting, the number of
\fakeTauh events is predicted by the weighted sum of the number of
events with one anti-isolated \tauh candidate minus the weighted number of
events with two anti-isolated \tauh candidates.

The validity of the extracted misidentification rates
is verified by defining additional \VRs
with either anti-isolated or isolated \tauh candidates,
mutually orthogonal to both the \SRs and \CRs.
The misidentification rates are used to normalize the
\fakeTauh samples from the anti-isolated \VRs to
a signal-depleted \VR with isolated \tauh candidates, where
the obtained background prediction is compared with the observed
data. This validation is performed separately for all channels and
data-taking periods.

In the \ltau channels, the \VRs are defined
by selecting events with the same selection criteria as those used
for the \SRs. However, instead of requiring the presence of one
\bjet\ in the \etau (\mtau) region, a low-\pt loosely identified and
loosely isolated displaced muon (electron) is required.
In the \ltautau channels, the validation of the background estimation is
performed using two \VRs. The first \VR is defined by using
similar event selections to the \SRs but vetoing events with
identified \bjets, while also removing any requirements on the \ST
variable.  For the second \VR, events are selected with identical
criteria as for the \SR, except that the \ST requirement is
inverted to satisfy $\ST<400\GeV$.
A summary of the event selection criteria for the \CRs and the
\VRs used is shown in \reftab{selections-CRs}, indicating only
selections that are different than their respective \SRs defined in
\reftab{selections-SRs}.

\begin{table}[!htb]
  \centering
  \topcaption
  {
    Offline selections applied to the reconstructed
    objects to obtain the \CRs\ and \VRs for the \fakeTauh candidate
    background estimation in the \ltau and \ltautau final
    states. Only differences with respect to the corresponding \SRs
    are shown. The \pt, \ptmiss, and \ST variables are reported in
    units of \GeVns, and \charge{} in units of $e$. The symbol $\star$
    is used to represent an electron (muon) for the \etau (\mtau)
    final states, and a \tauh object in the \etautau and
    \mtautau final states.
  }
  \label{tab:selections-CRs}
  \begin{tabular}{l l l l l l l l}
    \multicolumn{1}{c}{Object} & \multicolumn{1}{c}{Selection}     & \multicolumn{2}{c}{Control Regions} & \multicolumn{4}{c}{Validation Regions}                                                                                                 \\
                               &                                   & \multicolumn{1}{c}{\ltau}           & \multicolumn{1}{c}{\ltautau}           & \multicolumn{1}{c}{\etau} & \multicolumn{1}{c}{\mtau} & \multicolumn{2}{c}{\ltautau}          \\
    \hline
    \multirow{4}{*}{Displaced electrons}
                               & $N$                               & \NA                                 & \NA                                    & \NA                       & $=$1                      & \NA                          & \NA    \\
                               & $\pt$                             & \NA                                 & \NA                                    & \NA                       & $>$10                    & \NA                          & \NA    \\
                               & $\abseta$                         & \NA                                 & \NA                                    & \NA                       & $<$2.5                   & \NA                          & \NA    \\
                               & $\rIso{}$                         & \NA                                 & \NA                                    & \NA                       & $<$0.25                  & \NA                          & \NA    \\
    \\
    \multirow{4}{*}{Displaced muons}
                               & $N$                               & \NA                                 & \NA                                    & $=$1                      & \NA                       & \NA                          & \NA    \\
                               & $\pt$                             & \NA                                 & \NA                                    & $>$10                    & \NA                       & \NA                          & \NA    \\
                               & $\abseta$                         & \NA                                 & \NA                                    & $<$2.4                   & \NA                       & \NA                          & \NA    \\
                               & $\rIso{}$                         & \NA                                 & \NA                                    & $<$0.25                  & \NA                       & \NA                          & \NA    \\
    \\
    \tauh objects
                               & $\abs{\charge{\ltauhORtauhtauh}}$ & \NA                                 & $=$2                                   & \NA                       & \NA                       & \NA                          & \NA    \\
    \\
    \bjets                     & $N$                               & \NA                                 & \NA                                    & $\geq$0                  & $\geq$0                  & $=$0                         & \NA    \\
    \\
    \rTop                      & $N$                               & $=$1                               & \NA                                    & \NA                       & \NA                       & \NA                          & \NA    \\
    \\
    \multirow{2}{*}{Event-based}
                               & $\ptmiss$                         & $<$40                              & \NA                                    & \NA                       & \NA                       & \NA                          & \NA    \\
                               & $\ST$                             & \NA                                 & $>$0                                  & \NA                       & \NA                       & $>$0                        & $<400$ \\
    \hline
  \end{tabular}
\end{table}

\section{Search strategy}
\label{sec:strategy}
To maximize the sensitivity of the search, the discriminating
variable used to separate between the signal and background
processes is chosen separately for the \ltau and \ltautau final
states. For the latter, the search strategy focuses
on the reconstruction of the full \PHpm decay chain in order to search for
localized excesses in the \mT spectrum, as per
\refeq{mt}. For the hypothetical signal, the \mT distribution should
possess a Jacobian peak with an endpoint at $\mT=\mHpm$ that remains
unchanged by the transverse motion of the mother particle.
For the \ltau final states, the ambiguity in the
selection of the jets that go into the \mT calculation results in a
combinatorial background that smears the discriminating power of the
variable. Thus, while still valuable, the \mT is not
an optimal variable for signal extraction in the
\ltau channels.

In order to enhance the signal and background separation in the
\ltau final states, an \MVA \BDTG classifier is employed using
the \TMVA framework. The training was performed separately for each
simulated signal sample, final state, and data-taking period.
The inclusive \Vjets and \SMttbar simulated
samples were used to train against the background, weighted according
to their cross-sections. A total of 12 input variables were used for
the training of the \BDTG classifiers and include kinematic variables
of individual physics objects, as well as event-based variables.
They are summarized in \reftab{bdtg}.
All kinematic and event-based variable distributions were verified to
agree in shape between data and background estimates. 

\begin{table}[thb!]
  \centering
  \topcaption
  {
    The complete set of discriminating variables used in the
    training of the \BDTG classifier employed in the search strategy
    of the \ltau final states.
  }
  \label{tab:bdtg}
  \renewcommand{\arraystretch}{1.8}
  \begin{tabular}{l l}
    \multicolumn{1}{c}{Variable} & \multicolumn{1}{c}{Description}
    \\
    \hline
    \dPhi{\tauh}{\ptvecmiss}     & azimuthal angle between the \tauh and \ptvecmiss objects
    \\
    \dPhi{\ell}{\ptvecmiss}      & azimuthal angle between the $\ell$ and \ptvecmiss objects
    \\
    \PtjjPtHAsym                 & ratio of \pt sums calculated from $\ell$, \tauh, $j_{1}$, $j_{2}$ and \ptvecmiss
    \\
    \SumTwoJetOverHT             & ratio of \pt of the first two leading jets and the \HT
    \\
    \mtLtau                      & \mT reconstructed from $\ell$, \tauh, $j_{1}$, $j_{2}$, and \ptvecmiss
    \\
    \ThirdJetOverHT              & ratio of the \pt of the third leading jet and the \HT
    \\
    $\invMass{\ell}{\tauh}$      & invariant mass of the $\ell$ and \tauh objects
    \\
    \SumTwoJetAndLTOverHT        & ratio of \pt of first two leading jets plus \LT and the \HT
    \\
    $\mT(\ell, \ptvecmiss)$      & \mT reconstructed from the $\ell$ and \ptvecmiss objects
    \\
    $\pt^{\tauh}$                & transverse momentum of \tauh object
    \\
    $N_{\text{jets}}$            & number of selected jets in the event
    \\
    $N_{\rTop}$                  & number of selected \rTop objects in the event
    \\
    \hline
  \end{tabular}
\end{table}
\renewcommand{\arraystretch}{1.2}

\begin{figure}[!thb]
  \centering
  \includegraphics[width=0.49\textwidth]{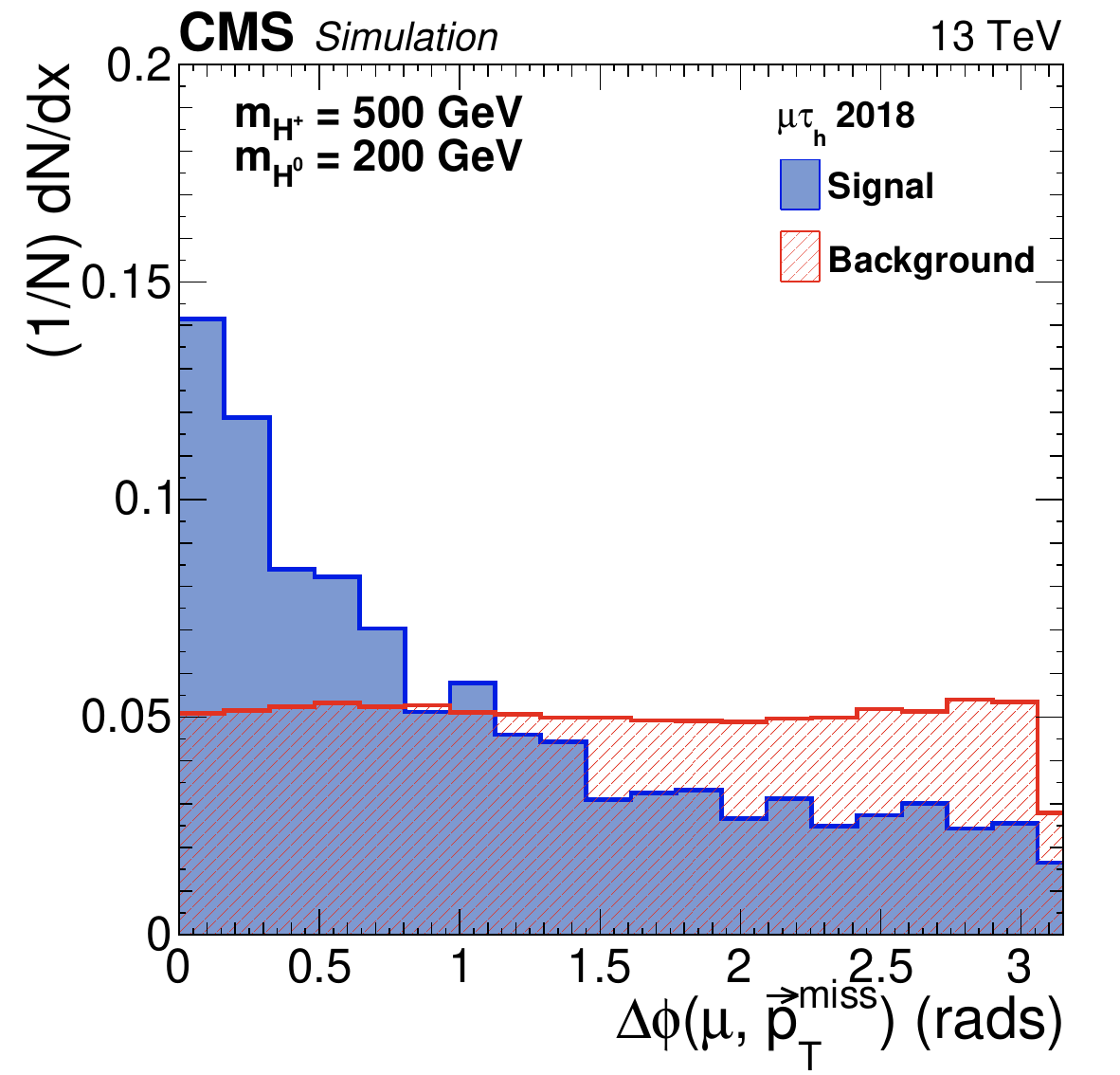}
  \includegraphics[width=0.49\textwidth]{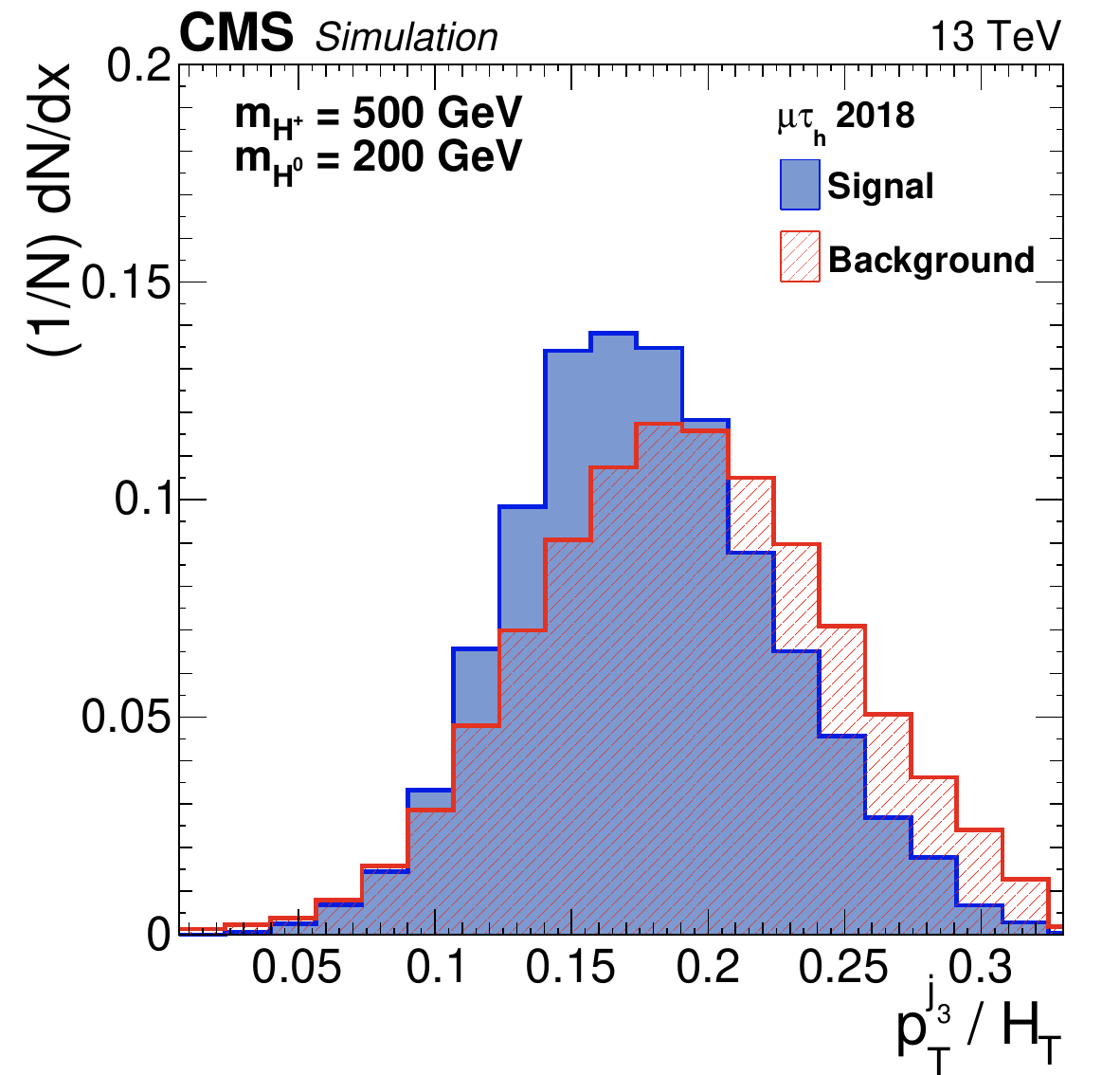}
  \includegraphics[width=0.49\textwidth]{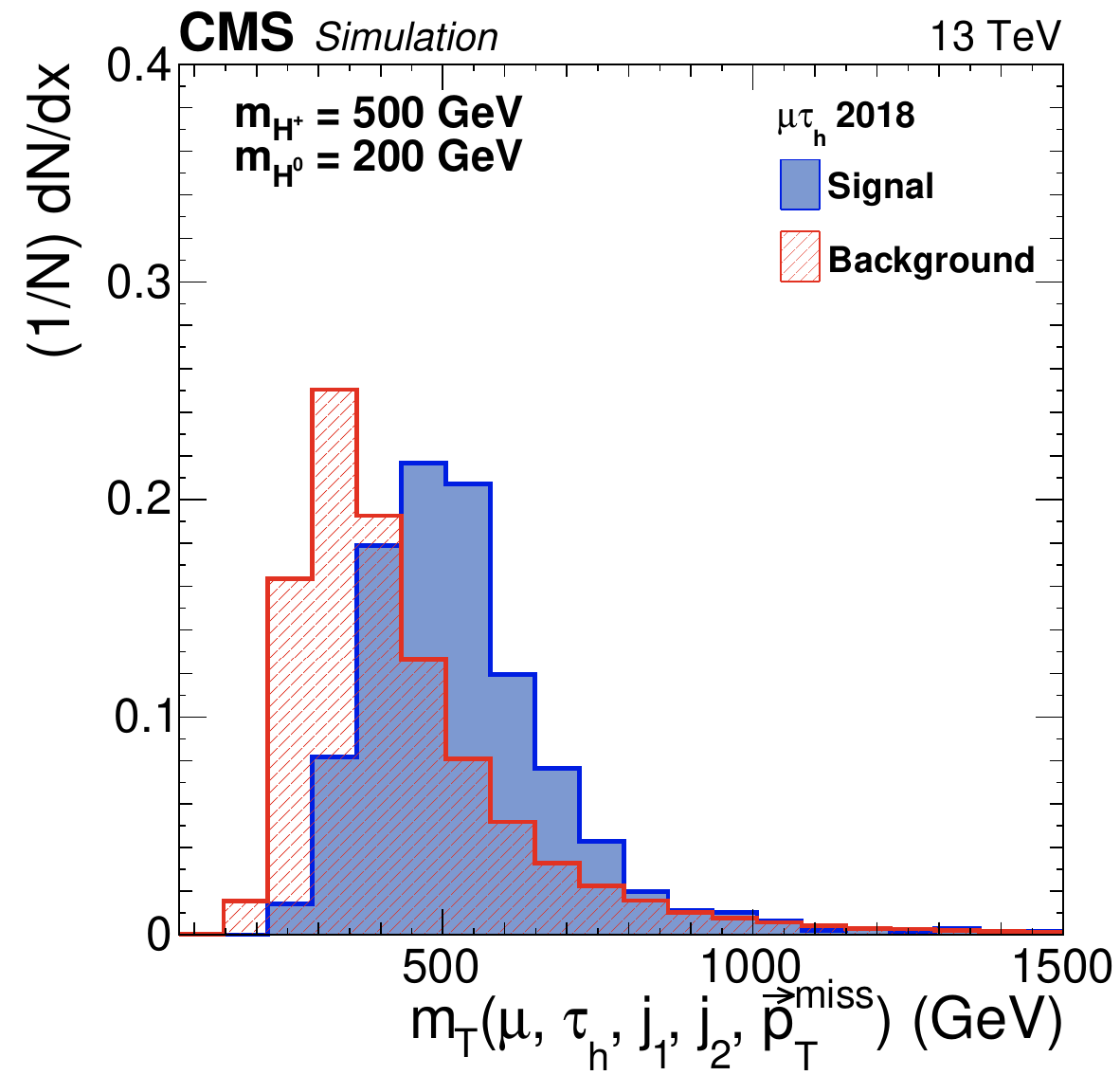}
  \caption{
    Three of the \BDTG input variables used for the \mtau final
    state, assuming a signal with mass \mSignal\ and 2018 data-taking
    conditions: the azimuthal angle between the \PGm and \ptvecmiss
    objects (\cmsTopLeft), the ratio of the \pt of the third leading jet
    and the \HT (\cmsTopRight), and the transverse mass reconstructed from
    the \PGm, \tauh, $j_{1}$, $j_{2}$, and \ptvecmiss objects (\cmsBottom).
    Both signal and background distributions are normalized to unit area.
  }
  \label{fig:bdtg-variables-top3}
\end{figure}

Three of these variables are shown in
\reffig{bdtg-variables-top3} for the \mtau final state and the 2018
data-taking period. Figure~\ref{fig:bdtg-variables-top3} (\cmsTopLeft)
shows the azimuthal angle between the \PGm and \ptvecmiss, denoted
as \dPhi{\PGm}{\ptvecmiss}. In the same figure (\cmsTopRight) the
ratio of the \pt of the third leading jet and the \HT, denoted as
\ThirdJetOverHToneline is shown.
Finally, the transverse mass \mtMuTau reconstructed from the selected
\PGm, \tauh, $j_{1}$, $j_{2}$, and \ptvecmiss objects is shown in
\reffig{bdtg-variables-top3} (\cmsBottom).
Instead of being used as an input to the \BDTG, the event-variable
\charge{\ltau} is used for the categorization of \ltau events, as it
increases signal sensitivity over the entire mass spectrum.
In particular, the \sSign selection significantly
suppresses the \SMttbar contribution that dominates the \oSign
category, leaving the misidentified \tauh as the dominant background.

\section{Systematic uncertainties}
\label{sec:systematics}

The systematic uncertainties from various experimental
and theoretical sources can affect the expected event yield (rate
uncertainties), the shape of the fit discriminant (shape
uncertainties), or both. Log-normal (Gaussian) a priori distributions
are assumed for rate (shape) uncertainties.
Partial and complete correlations between the
uncertainties in different categories and years are taken into account,
depending on the way they are derived. All experimental
sources are treated as correlated across categories but as uncorrelated
across years, unless otherwise specified. All theoretical sources are
treated as correlated across all categories and years.
A summary of all sources of systematic uncertainties discussed in
this section is given in \reftab{systematics}.

Apart from the various experimental uncertainties, the statistical
analysis of the results employs an uncertainty model
that also accounts for uncertainties due to the limited population of
template distributions in signal and background modeling.
These statistical uncertainties can lead to fluctuations in nominal
predictions and their effect is individually incorporated for each
template bin with the \textit{Barlow--Beeston lite}
approach~\cite{Barlow:1993dm,Conway:2011in}.
Each bin is assigned a combined statistical uncertainty,
and these uncertainties are treated
as uncorrelated among other bins, categories, channels, and data sets.

\subsection{Experimental sources}

The integrated luminosity for each year of data taking
is measured individually with an uncertainty in the 1.2--2.5\%
range~\cite{CMS-LUM-17-003,CMS-PAS-LUM-17-004,CMS-PAS-LUM-18-002}.
The total integrated luminosity for the period 2016--2018 has an
uncertainty of 1.6\%.  The improvement in precision reflects the
uncorrelated time evolution of some systematic effects.
These effects are applied as rate uncertainties to all simulated
processes and thus only affect the expected yield of events, but not
the individual shapes of the fit discriminants.

Uncertainties related to the electron and muon trigger efficiencies
arise from the fact that simulated events are corrected to match the
efficiencies measured in data. The uncertainties in the corrections,
which depend on the \pt and $\eta$ of the trigger object, lead to
rate and shape effects in the fit discriminants that
amount to 1--4\%.

During the 2016--2017 data taking, a gradual shift in the timing of
the \ECAL\ inputs to the first-level trigger in the region of $\abseta > 2.0$
caused a trigger inefficiency of
$\approx$10--20\% for events containing an electron (jet)
with $\pt \gtrsim 50$ (100)\GeV and in the
region $2.5 < \abs{\eta} < 3.0$, depending on \pt, $\eta$, and time.
Corresponding correction factors were derived from data and applied to the
acceptance evaluated from simulation. The related uncertainties, which
are treated as correlated between the two years and all categories,
are found to affect the expected event yields by $\approx$1\%
when propagated to the final fit discriminants.

The uncertainty due to the pileup modeling in simulated
samples is estimated by varying the total inelastic \pp cross section
used to estimate the number of pileup events in data. The nominal
value of 69.2\unit{mb} is varied by
5\%~\cite{ATLAS:2016pu,Sirunyan:2018nqx} and the effect is propagated
through all event selections. The resulting uncertainty amounts to up
to 5.5\%, and is treated as correlated across all categories and  years.

Uncertainties associated with the identification efficiency
for electrons and muons are propagated as variations to the final fit discriminants.
They are treated as correlated across all categories and
years and result in shape-altering variations that change the total
event yield by about 0.1--2.4\%. The uncertainty related to the
vetoing of leptons passing loose selection criteria is between 0.1 and 2.0\%.

The uncertainties associated with the \tauh identification efficiency are
evaluated in \pt bins of the \tauh object and range between
0.1 and 5.3\%. The \TES uncertainties are found to be up to about 2.0\%,
depending on the decay mode of the \tauh object. For the energy and
momentum scale of electrons and muons misidentified as \tauh candidates,
the relevant corrections depend on the \pt and decay mode of the
candidate and their uncertainties are of \orderOf{1\%}. The
above-mentioned uncertainties lead to shape effects and are treated as
uncorrelated across $\pt$ bins and decay modes.

The \JES uncertainties are specified as functions of
jet \pt and $\eta$ and are treated as correlated across all categories and
years. They are estimated by shifting the energy of jets
and propagating these shifts through the analysis selections. This results
in rate- and shape-altering variations of
\orderOf{5\%}. The energy resolution of simulated jets is adjusted to match the
resolution observed in data. The uncertainties in the \JER are
evaluated by smearing the jet energies around their nominal values.
These are treated as uncorrelated across the years
and correlated across categories and result in an overall effect of
\orderOf{5\%}.

Jet energy uncertainties are also propagated to the \ptvecmiss
calculation to account for the fact that it primarily relies on the accurate
measurement of the reconstructed physics objects. Another uncertainty
in the \ptvecmiss measurement is related to the unclustered energy in the
event. It refers to jets with $\pt < 10\GeV$ and \PF candidates not
clustered into jets. This \UES uncertainty impacts both the rate and
the shape of the fit discriminants and has an overall effect of up to 2.4\%.

The efficiency of classifying jets as b tagged is different in data
than in simulation. To correct for this effect
\pt-dependent corrections are incorporated. The systematic
uncertainties in these tagging and mistagging efficiency corrections are
treated as rate and shape altering. They are found to have an
effect of \orderOf{5\%} in the expected event yields.

The uncertainties associated with the \rTop tagging and mistagging
efficiency corrections are only relevant for the \ltau categories.
They are evaluated in bins of \pt of the selected
\rTop and the various sources of uncertainty are treated as uncorrelated.
These include effects due to the generator-level matching definition,
the damping of radiation with high \pt,
the modeling of the first emission, the scale radiation, the color
reconnection strength, the assumed value of the top quark mass, and
the tuning of the \UE parameters. These are propagated to the fit
discriminants of the \ltau final states as shape uncertainties and
are found to change the total event yield by about 5\% for tagging and
8\% for mistagging.

For the \fakeTauh background estimation three distinct components of
uncertainties are defined. The first component is the statistical
uncertainty associated with the evaluation of the misidentification
rates in the dedicated \CRs. The various parametrization bins
that are used in the \ltau and \ltautau channels for this measurement,
as described in \refsec{background}, are treated as uncorrelated.
The propagation of the statistical uncertainty in the final fit
discriminants has an overall effect of \orderOf{10\%}.
The second component is concerned with a rate and shape-altering
uncertainties implemented to address the level of agreement of closure
tests in the \VRs. Their impact on the expected event yield is \orderOf{10\%}.
The third component  accounts for the difference in sample composition
between the \CRs in which the misidentification rates are
determined, and the \SRs in which they are applied.
It is treated as uncorrelated between the parametrization
bins and when propagated to the final fit discriminants brings about
an overall effect of up to 18\%.

\subsection{Theoretical sources}
The systematic uncertainties related to theoretical considerations
mainly arise due to missing higher-order \QCD corrections
and uncertainties in the \PDF sets. An additional source of uncertainty
concerns the assumed values of the top quark
mass \mtop and that of the strong coupling \alpS in parton
showers~\cite{Heinemeyer:2013tqa}. These affect both the total and
differential cross sections of the processes, yielding uncertainties
on the overall normalization of the simulated processes and the
acceptance of the event selection. All effects are taken into account
as rate uncertainties.

For the \ttbar and \SingleTop processes,
the effect of \mtop on the cross sections is evaluated
by varying its nominal value of 172.5\GeV by 1.0\GeV.
The effects from the \RF scales on the acceptance and
cross sections are evaluated by varying them independently by factors
of one-half and two with respect to their nominal values,
respectively. Extreme variations where one scale is varied by one-half
and the other one by two are excluded. The effect on the event yield
from simulated events is then calculated by enveloping the maximum
variation with respect to  the nominal fit discriminants, as
recommended in \refcite{deFlorian:2016spz}.

The \PDF uncertainties are also treated as fully correlated for all
processes and categories. They are also correlated between simulated
samples that share the same dominant partons in the initial state of
the \ME calculations~\cite{Butterworth:2015oua}.

\begin{table}[!htb]
  \centering
  \topcaption{
    Summary of all sources of systematic uncertainties discussed in the
    text.  The first column identifies the source of uncertainty and,
    where applicable, the process that it applies to.
    The second column indicates with a check mark $\checkmark$ or dash
    \NA\ whether or not the nuisance parameter also affects
    the shape of the fit discriminant. The third column, which is
    subdivided into four event categories, presents the percentage \% impact of these
    nuisance parameters on the expected event yields, before
    simultaneous fitting the data for the background-only hypothesis.
    A range of such values represents the minimum and maximum values observed
    through the different samples and data eras, with apparent disparities
    also attributed to the limited sample size of minor backgrounds.
    The last two columns indicate whether or not the nuisance parameters
    are correlated across years and categories.
    A dagger $\dagger$ designates that a nuisance parameter is only
    partially correlated across years or categories.
  }
  \label{tab:systematics}
  \cmsTable{
    \begin{tabular}{l c c c c c c c}
                                                &              & \multicolumn{4}{c}{Category}         & \multicolumn{2}{c}{Correlated across}                                                         \\
      Uncertainty source                        & Shape        & \etau                                & \mtau                                 & \etautau     & \mtautau & Years        & Categories   \\
      \hline
      \multicolumn{8}{c}{Experimental}                                                                                                                                                                \\
      Integrated luminosity                     & \NA          & \multicolumn{4}{c}{1.2--2.5}         & $\checkmark^{\dagger}$                & $\checkmark$                                          \\
      Trigger efficiency                        & $\checkmark$ & 0.9--4.2                             & 0.5--2.9                              & 1.2--3.0     & 0.1--0.3 & \NA          & $\checkmark$ \\
      Trigger timing inefficiency               & $\checkmark$ & \multicolumn{2}{c}{\NA}              & 0.1--0.3                              & 0.1--0.3     & \NA      & $\checkmark$                \\
      Pileup                                    & $\checkmark$ & 0.2--2.9                             & 0.1--1.6                              & 0.1--5.5     & 0.1--2.3 & $\checkmark$ & $\checkmark$ \\
      Electron identification                   & $\checkmark$ & 0.1--2.4                             & \NA                                   & 0.1--1.9     & \NA      & $\checkmark$ & $\checkmark$ \\
      Muon identification                       & $\checkmark$ & \NA                                  & 0.4--1.6                              & \NA          & 0.1--1.1 & $\checkmark$ & $\checkmark$ \\
      Lepton veto                               & \NA          & 0.1--2.0                             & 0.1--2.0                              & 0.1--2.0     & 0.1--2.0 & \NA          & $\checkmark$ \\
      \tauh identification                      & $\checkmark$ & 0.1--3.6                             & 0.1--3.2                              & 4.2--5.2     & 4.3--5.3 & \NA          & $\checkmark$ \\
      \TES                                      & $\checkmark$ & 0.2--1.8                             & 0.2--0.4                              & 0.8--1.5     & 0.8--1.2 & \NA          & $\checkmark$ \\
      \EToTauh misidentification                & $\checkmark$ & 0.2--1.5                             & 0.1--0.4                              & 0.3--0.6     & 0.2--0.6 & \NA          & $\checkmark$ \\
      \MToTauh misidentification                & $\checkmark$ & 0.1--1.6                             & 0.1--0.3                              & 0.1--0.3     & 0.1--0.3 & \NA          & $\checkmark$ \\
      Jet energy scale                          & $\checkmark$ & 1.1--4.9                             & 1.2--4.2                              & 1.6--3.6     & 1.6--2.4 & $\checkmark$ & $\checkmark$ \\
      Jet energy resolution                     & $\checkmark$ & 0.3--3.1                             & 0.3--3.1                              & 1.1--4.6     & 1.2--3.1 & \NA          & $\checkmark$ \\
      \bjet identification                      & $\checkmark$ & 2.5--5.4                             & 2.5--5.1                              & 2.4--4.2     & 3.2--4.2 & \NA          & $\checkmark$ \\
      \bjet misidentification                   & $\checkmark$ & 2.4--4.1                             & 2.2--4.5                              & 1.0--2.6     & 1.7--2.5 & \NA          & $\checkmark$ \\
      Unclustered energy scale                  & $\checkmark$ & 0.1--1.9                             & 0.2--1.5                              & 0.5--1.7     & 0.3--2.4 & \NA          & $\checkmark$ \\
      \rTop tagging                             & $\checkmark$ & 1.5--7.6                             & 1.3--7.5                              & \NA          & \NA      & \NA          & $\checkmark$ \\
      \rTop mistagging                          & $\checkmark$ & 1.7--4.9                             & 1.7--5.4                              & \NA          & \NA      & \NA          & $\checkmark$ \\
      \JetToTauh misidentification              & $\checkmark$ & 17.8--21.1                           & 18.2--22.4                            & 14.8         & 10.5     & \NA          & \NA          \\
      \multicolumn{8}{c}{Theoretical}                                                                                                                                                                 \\
      Top quark mass ($\SMttbar$)               & \NA          & \multicolumn{4}{c}{2.2}              & $\checkmark$                          & $\checkmark$                                          \\
      Top quark mass (\SingleTop)               & \NA          & \multicolumn{4}{c}{2.8}              & $\checkmark$                          & $\checkmark$                                          \\
      Acceptance \PHpm (\RF scale, PDF)         & \NA          & \multicolumn{4}{c}{5.3}              & $\checkmark$                          & $\checkmark$                                          \\
      Acceptance \SMttbar (\RF scale, PDF)      & \NA          & \multicolumn{4}{c}{$-2.8$ to $+2.0$} & $\checkmark$                          & $\checkmark$                                          \\
      Acceptance \SingleTop (\RF scale, PDF)    & \NA          & \multicolumn{4}{c}{$-2.0$ to $+0.3$} & $\checkmark$                          & $\checkmark$                                          \\
      Acceptance \ttX (\RF scale, PDF)          & \NA          & \multicolumn{4}{c}{2.0}              & $\checkmark$                          & $\checkmark$                                          \\
      Acceptance EW (\RF scale, PDF)            & \NA          & \multicolumn{4}{c}{$<$1.0}           & $\checkmark$                          & $\checkmark$                                          \\
      Cross section \SMttbar (\RF scale, PDF)   & \NA          & \multicolumn{4}{c}{$-4.8$ to 5.5}    & $\checkmark$                          & $\checkmark$                                          \\
      Cross section \SingleTop (\RF scale, PDF) & \NA          & \multicolumn{4}{c}{5.3}              & $\checkmark$                          & $\checkmark$                                          \\
      Cross section \ttX (\RF scale, PDF)       & \NA          & \multicolumn{4}{c}{2.2}              & $\checkmark$                          & $\checkmark$                                          \\
      Cross section EW (\RF scale, PDF)         & \NA          & \multicolumn{4}{c}{5.4}              & $\checkmark$                          & $\checkmark$                                          \\
      \hline
    \end{tabular}
  }
\end{table}

\section{Results}
\label{sec:results}
Binned \MVA output distributions in the
\ltau analysis and \mT distributions in the
\ltautau analysis are used to test the compatibility of the observed
data with the presence or absence of a signal.
Data are split in 6 categories determined by the lepton flavor
in the final state (\Pe or \PGm), for each of the three years of data
taking. For the \ltauh final states the number of categories is double
due to the consideration of the sum of the electric charge of the
lepton and \tauh objects. Therefore there are 18 categories in total;
12 associated with the \ltau analysis and 6 with the \ltautau analysis.
A simultaneous binned maximum likelihood fit is performed over all
categories and data sets. The
likelihood incorporates all the systematic uncertainties described in
\refsec{systematics} as nuisance parameters, with shape variations
taken into account via continuous morphing~\cite{Conway:2011in}.

\begin{figure}[htbp!]
  \centering
  \includegraphics[width=0.49\textwidth]{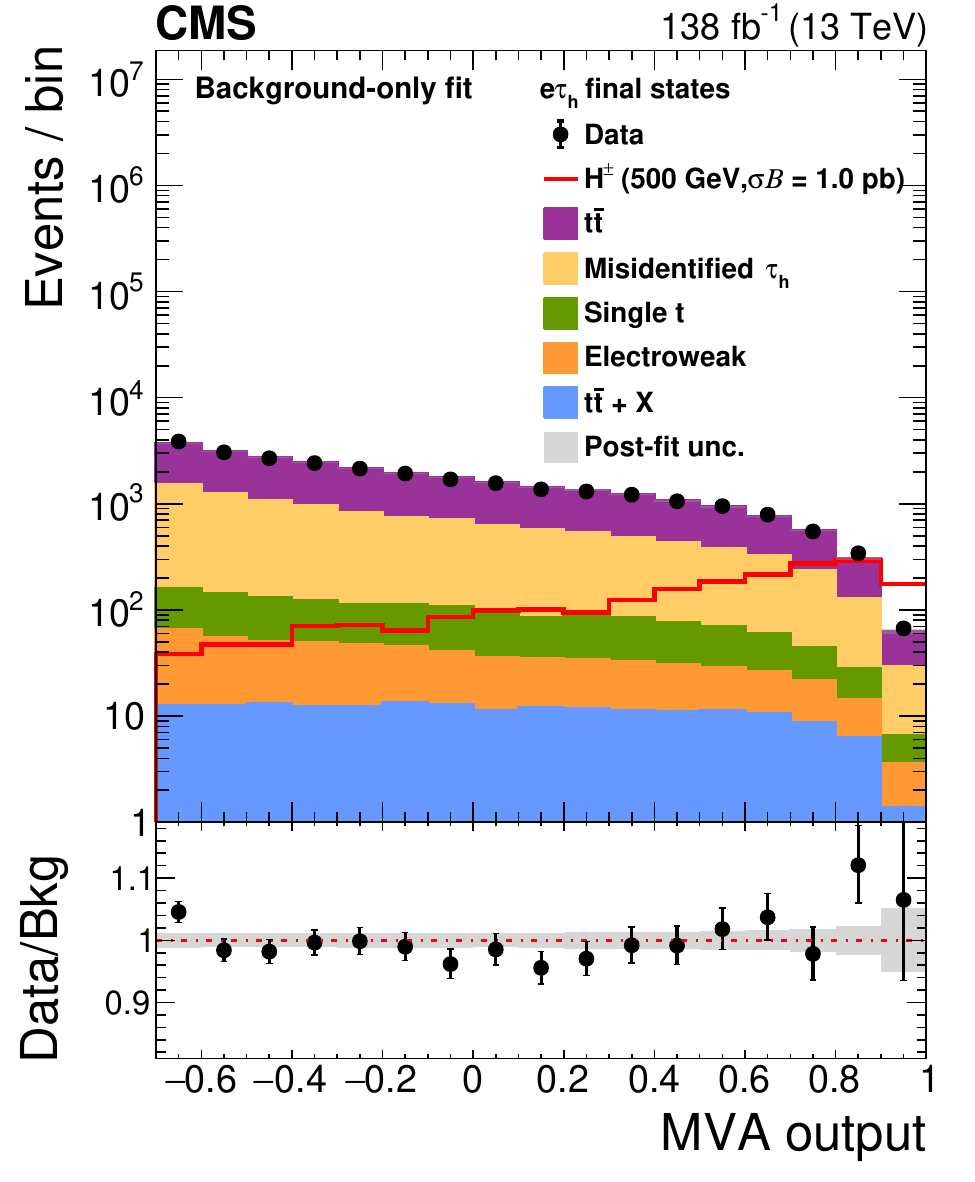}
  \includegraphics[width=0.49\textwidth]{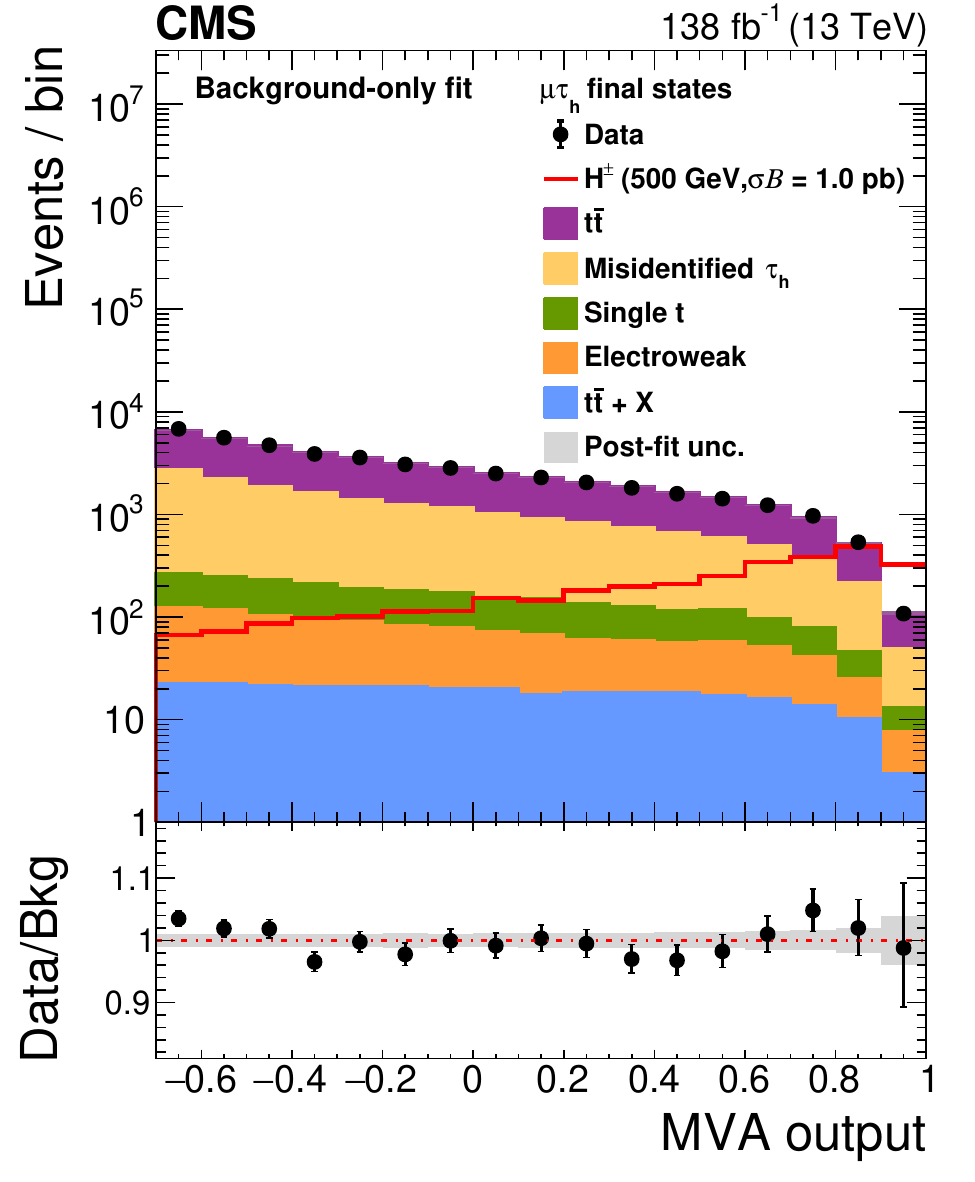}
  \caption{
    The \MVA output of the \BDTG for the \etau (left) and
    \mtau (right) final states used in the limit extraction, after
    a background-only fit to the data. The data sets of all
    categories have been added. The pre-fit contribution
    from \HpmToHW with masses \mSignal\ and $\mH = 200\GeV$ and
    $\xsDef = 1\unit{pb}$ is also shown.
  }
  \label{fig:postfit-ltau}
\end{figure}

No significant excess is found in any of the categories considered.
The distributions for the fit performed under the background-only
hypothesis are shown in \refdifig{postfit-ltau}{postfit-ltautau}
for the \ltau and \ltautau final states, respectively,
whereby all categories for each individual final state are
added into a single distribution. For the \ltautau final states, the
distributions are binned with variable width and according to the
statistical precision of the samples. In order to retain the shape of
the distributions, each bin is divided by their width.
The pre-fit contribution of a hypothetical \HpmToHW signal with
masses \mSignal\ and $\mH = 200\GeV$ is also shown, normalized assuming
that the product of the cross section and branching fraction
\xsDef is 1\unit{pb}. Tabulated results are provided in the
HEPData record for this
analysis~\cite{hepdata}.

\begin{figure}[htbp!]
  \centering
  \includegraphics[width=0.49\textwidth]{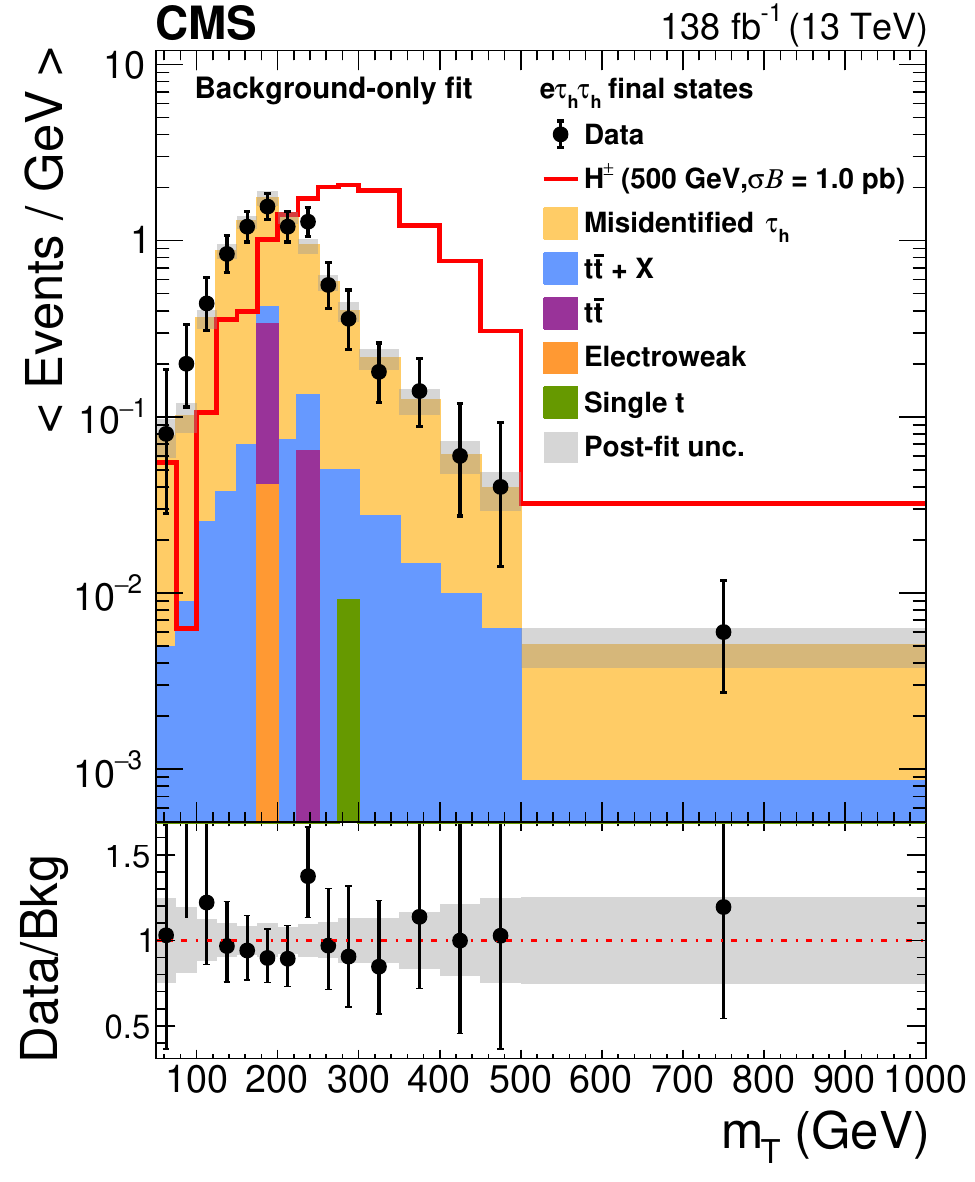}
  \includegraphics[width=0.49\textwidth]{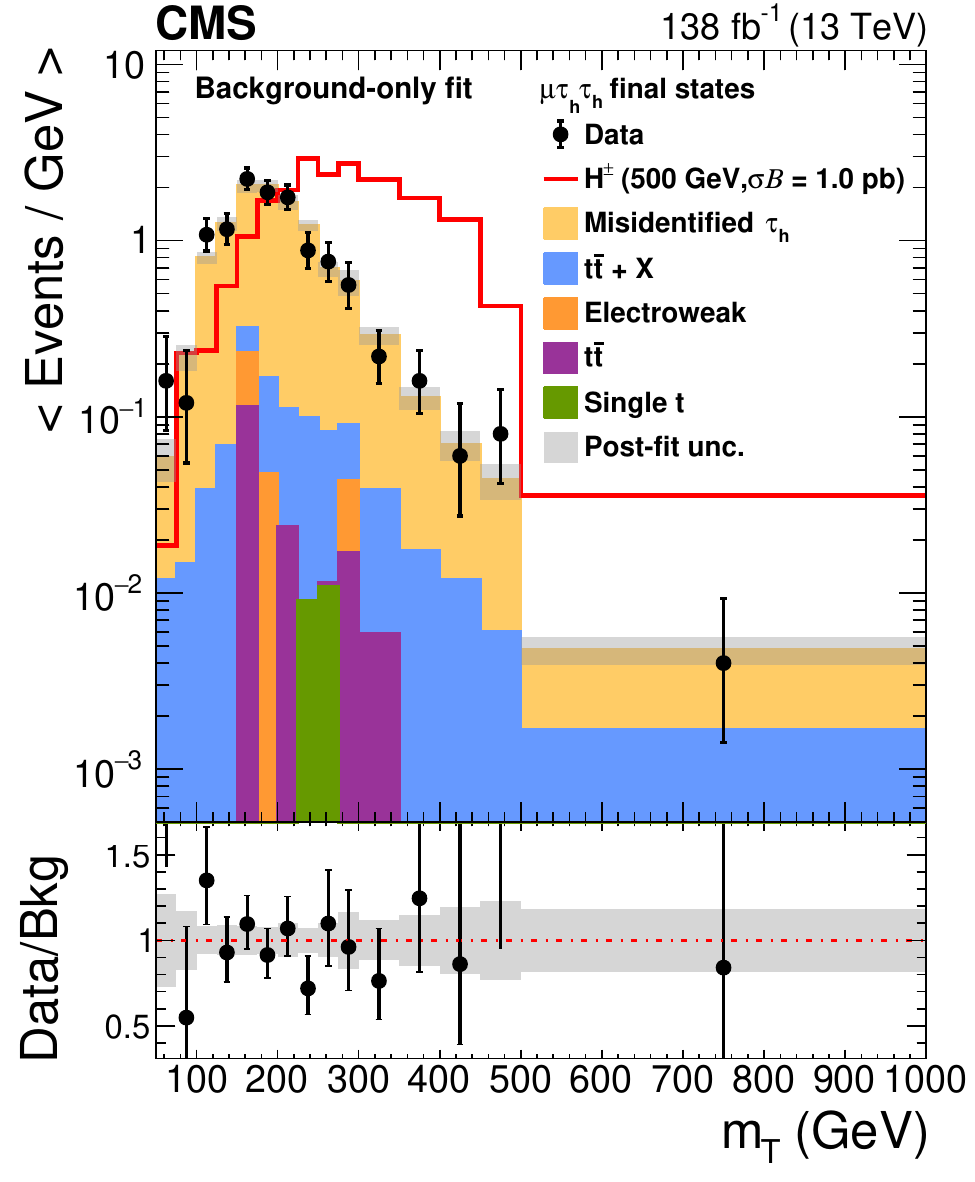}
  \caption{
    The \mT distributions for the \etautau (left) and
    \mtautau (right) final states used in the limit extraction, after
    a background-only fit to the data. The data sets of all categories
    have been added. The pre-fit contribution
    from \HpmToHW with masses \mSignal\ and $\mH = 200\GeV$
    and $\xsDef = 1\unit{pb}$ is also shown.
    The brackets $\langle\cdot\rangle$ signify that the plotted variable is averaged
    over an interval in which the event frequency may have changed
    considerably.
  }
  \label{fig:postfit-ltautau}
\end{figure}

\begin{figure}[ttb!]
  \centering
  \includegraphics[width=1.0\textwidth]{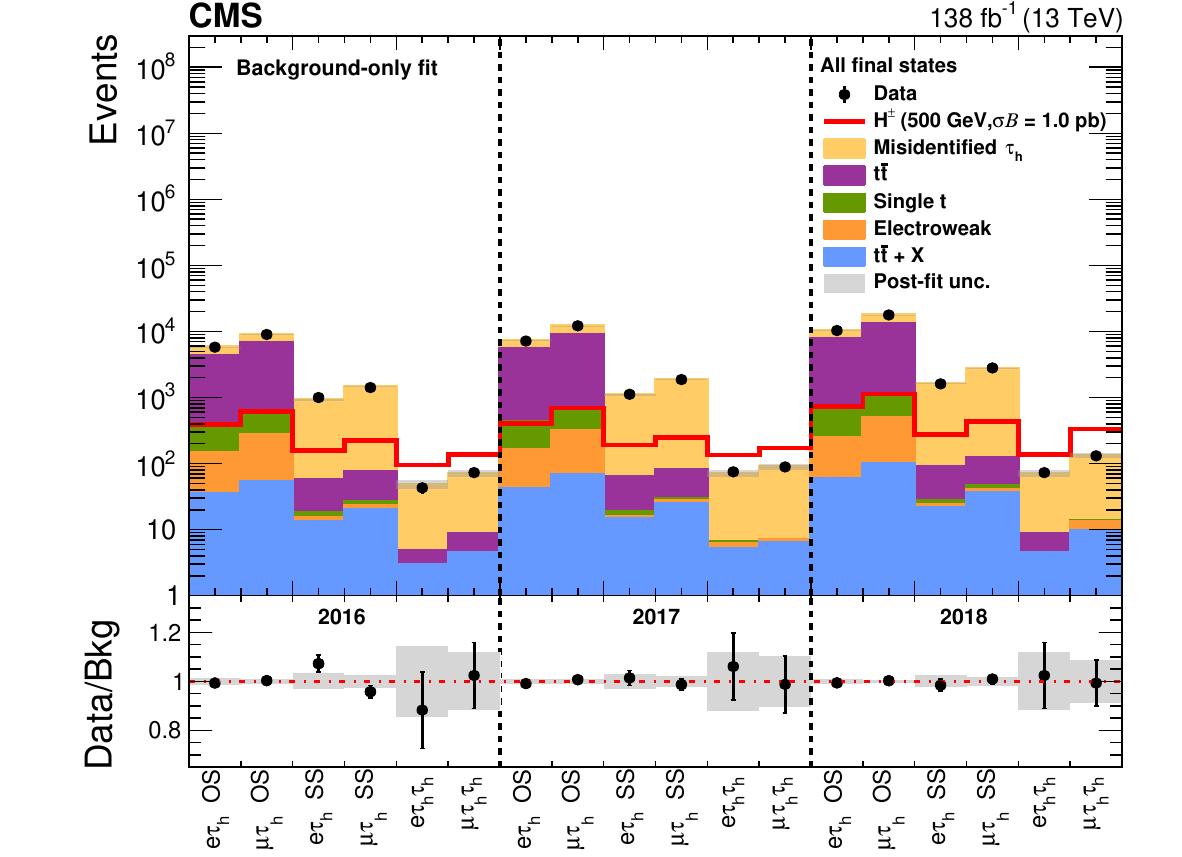}
  \caption{
    Observed event yields (black markers) for the 18 categories considered
    in this analysis, grouped into data sets that are represented by
    vertical dashed lines. The expected event yields (stacked
    histograms) resulting from a background-only fit to the data are
    also shown, broken down into various background processes.
    The solid red line represents the expected signal yields from
    \HpmToHW with masses \mSignal\ and $\mH = 200\GeV$, assuming $\xsDef = 1\unit{pb}$.
  }
  \label{fig:postfit-yields}
\end{figure}

The event rates, with the expected yields from the \SM backgrounds
normalized as resulting from a background-only fit to the data,
are shown in \reffig{postfit-yields}, for all data sets and
final states considered.

Upper limits on \xsDef
for a potential \PHpm signal are computed at
the 95\% \CL, using the modified frequentist \CLs
criterion~\cite{Junk:1999kv,Read:2002hq}. The definition of the
profile likelihood test statistic is as defined in
\refcite{Chatrchyan:2012tx}, using the asymptotic
approximation~\cite{Cowan:2010js}.

\begin{figure}[htbp!]
  \centering
  \includegraphics[width=0.49\textwidth]{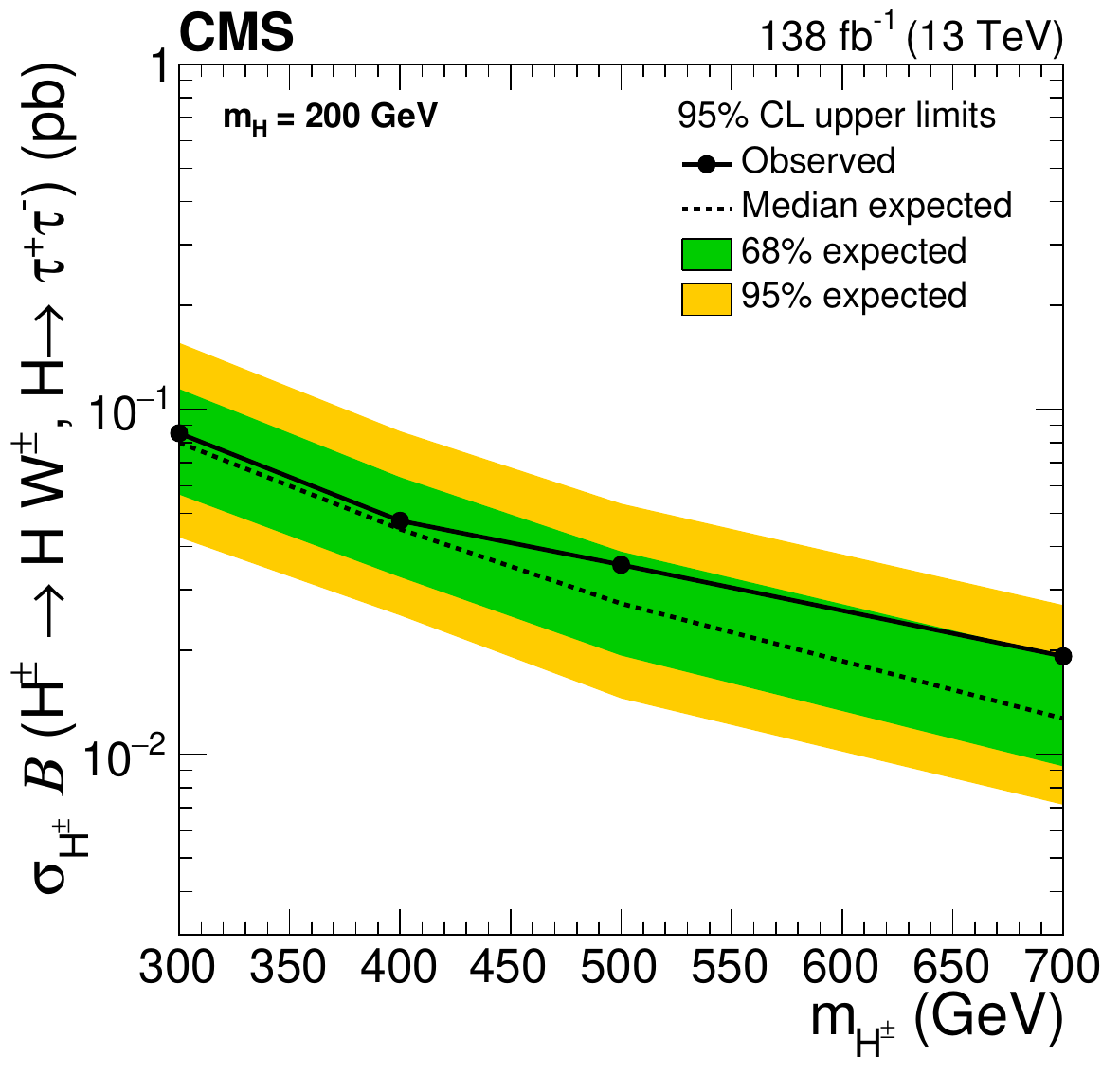}
  \includegraphics[width=0.49\textwidth]{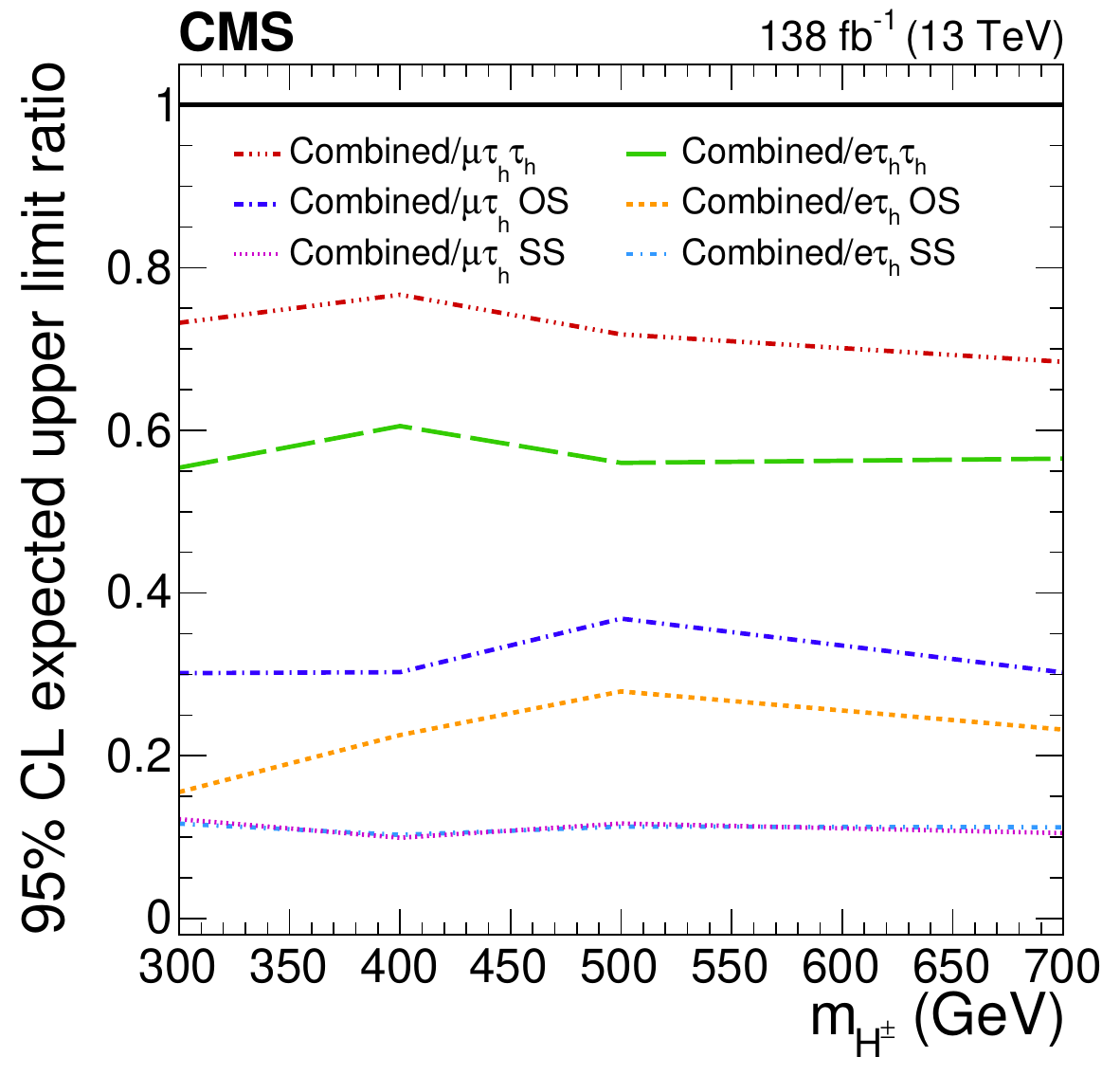}
  \caption{
    Expected and observed upper limits at 95\% \CL on the product of
    cross section and branching fraction \xsDef as a function of
    \mHpm and assuming $\mH = 200\GeV$
    for the combination of all final states considered
    (\cmsLeft).
    The observed upper limits are represented by a
    solid black line and circle markers. The median
    expected limit (dashed line), 68\% (inner green band), and 95\%
    (outer yellow band) confidence intervals are also shown.
    The relative expected contributions of each
    final state to the overall combination are also presented (\cmsRight).
    The black solid line corresponds to the combined expected limit,
    while the red dash-dotted, green dashed, blue dashed-dotted, 
    orange dashed, magenta dotted, and light blue dash-dotted lines
    represent the relative contributions from the \mtautau, \etautau,
    \mtau\ OS, \etau\ OS, \mtau\ SS, and \etau\ SS categories, respectively. 
  }
  \label{fig:limits}
\end{figure}

The upper limit with all final states, categories,
and years combined is shown in \reffig{limits} (\cmsLeft).
The observed upper limit on \xsDef varies between \xsLimit. In the
same figure (\cmsRight), the expected sensitivity from each
contributing final state is also shown. The \ltautau final states are
the most sensitive in the whole \mHpm range from \massRange, while the
\ltau final states improve the overall sensitivity by 10--35\%. 
The dependence of these results on the assumed \PH mass was studied
with full simulation for two additional mass points, $\mH= 125\GeV$
and $\mH= 150\GeV$.  The rate and shape of the transverse mass of the charged
Higgs boson in \refeq{mt}, which is the fit discriminant of the most sensitive  
categories, \etautau\ and \mtautau, were shown to be unchanged. As a
result, the observed upper limits are not expected to show a strong
dependence on small variations to the mass of the heavy neutral Higgs
boson.

\clearpage
\section{Summary}
\label{sec:summary}
Results are presented from a search for a charged Higgs boson
\PHpm decaying into a heavy neutral Higgs boson \PH and a \PW
boson. Events are selected with exactly one isolated electron or muon,
targeting event topologies whereby the \PH decays
into a pair of tau leptons with at least one decaying hadronically (\tauh).
Four distinct final states are considered: \myFS.
The analysis uses proton-proton collision data recorded by the CMS
detector at $\sqrt{s}=13\TeV$, corresponding to an integrated
luminosity of \intLumi. No significant deviation is observed from
standard model expectations. Upper limits at 95\% confidence level
are set on the product of the cross section and branching fraction for
an \PHpm in the mass range of 300--700\GeV,
assuming an \PH with a mass of 200\GeV. The
observed limits range from 0.085\unit{pb} for an
\PHpm mass of 300\GeV to 0.019\unit{pb}
for a mass of 700\GeV. These are the first limits on
\PHpm production in the $\PHpm \to \PH \PWpm$ decay
channel at the LHC.

\begin{acknowledgments}
We congratulate our colleagues in the CERN accelerator departments for the excellent performance of the LHC and thank the technical and administrative staffs at CERN and at other CMS institutes for their contributions to the success of the CMS effort. In addition, we gratefully acknowledge the computing centers and personnel of the Worldwide LHC Computing Grid and other centers for delivering so effectively the computing infrastructure essential to our analyses. Finally, we acknowledge the enduring support for the construction and operation of the LHC, the CMS detector, and the supporting computing infrastructure provided by the following funding agencies: BMBWF and FWF (Austria); FNRS and FWO (Belgium); CNPq, CAPES, FAPERJ, FAPERGS, and FAPESP (Brazil); MES and BNSF (Bulgaria); CERN; CAS, MoST, and NSFC (China); MINCIENCIAS (Colombia); MSES and CSF (Croatia); RIF (Cyprus); SENESCYT (Ecuador); MoER, ERC PUT and ERDF (Estonia); Academy of Finland, MEC, and HIP (Finland); CEA and CNRS/IN2P3 (France); BMBF, DFG, and HGF (Germany); GSRI (Greece); NKFIH (Hungary); DAE and DST (India); IPM (Iran); SFI (Ireland); INFN (Italy); MSIP and NRF (Republic of Korea); MES (Latvia); LAS (Lithuania); MOE and UM (Malaysia); BUAP, CINVESTAV, CONACYT, LNS, SEP, and UASLP-FAI (Mexico); MOS (Montenegro); MBIE (New Zealand); PAEC (Pakistan); MES and NSC (Poland); FCT (Portugal); MESTD (Serbia); MCIN/AEI and PCTI (Spain); MOSTR (Sri Lanka); Swiss Funding Agencies (Switzerland); MST (Taipei); MHESI and NSTDA (Thailand); TUBITAK and TENMAK (Turkey); NASU (Ukraine); STFC (United Kingdom); DOE and NSF (USA).

\hyphenation{Rachada-pisek} Individuals have received support from the Marie-Curie program and the European Research Council and Horizon 2020 Grant, contract Nos.\ 675440, 724704, 752730, 758316, 765710, 824093, 884104, and COST Action CA16108 (European Union); the Leventis Foundation; the Alfred P.\ Sloan Foundation; the Alexander von Humboldt Foundation; the Belgian Federal Science Policy Office; the Fonds pour la Formation \`a la Recherche dans l'Industrie et dans l'Agriculture (FRIA-Belgium); the Agentschap voor Innovatie door Wetenschap en Technologie (IWT-Belgium); the F.R.S.-FNRS and FWO (Belgium) under the ``Excellence of Science -- EOS" -- be.h project n.\ 30820817; the Beijing Municipal Science \& Technology Commission, No. Z191100007219010; the Ministry of Education, Youth and Sports (MEYS) of the Czech Republic; the Hellenic Foundation for Research and Innovation (HFRI), Project Number 2288 (Greece); the Deutsche Forschungsgemeinschaft (DFG), under Germany's Excellence Strategy -- EXC 2121 ``Quantum Universe" -- 390833306, and under project number 400140256 - GRK2497; the Hungarian Academy of Sciences, the New National Excellence Program - \'UNKP, the NKFIH research grants K 124845, K 124850, K 128713, K 128786, K 129058, K 131991, K 133046, K 138136, K 143460, K 143477, 2020-2.2.1-ED-2021-00181, and TKP2021-NKTA-64 (Hungary); the Council of Science and Industrial Research, India; the Latvian Council of Science; the Ministry of Education and Science, project no. 2022/WK/14, and the National Science Center, contracts Opus 2021/41/B/ST2/01369 and 2021/43/B/ST2/01552 (Poland); the Funda\c{c}\~ao para a Ci\^encia e a Tecnologia, grant CEECIND/01334/2018 (Portugal); the National Priorities Research Program by Qatar National Research Fund; MCIN/AEI/10.13039/501100011033, ERDF ``a way of making Europe", and the Programa Estatal de Fomento de la Investigaci{\'o}n Cient{\'i}fica y T{\'e}cnica de Excelencia Mar\'{\i}a de Maeztu, grant MDM-2017-0765 and Programa Severo Ochoa del Principado de Asturias (Spain); the Chulalongkorn Academic into Its 2nd Century Project Advancement Project, and the National Science, Research and Innovation Fund via the Program Management Unit for Human Resources \& Institutional Development, Research and Innovation, grant B05F650021 (Thailand); the Kavli Foundation; the Nvidia Corporation; the SuperMicro Corporation; the Welch Foundation, contract C-1845; and the Weston Havens Foundation (USA).
\end{acknowledgments}

\bibliography{auto_generated}
\cleardoublepage \appendix\section{The CMS Collaboration \label{app:collab}}\begin{sloppypar}\hyphenpenalty=5000\widowpenalty=500\clubpenalty=5000
\cmsinstitute{Yerevan Physics Institute, Yerevan, Armenia}
{\tolerance=6000
A.~Tumasyan\cmsAuthorMark{1}\cmsorcid{0009-0000-0684-6742}
\par}
\cmsinstitute{Institut f\"{u}r Hochenergiephysik, Vienna, Austria}
{\tolerance=6000
W.~Adam\cmsorcid{0000-0001-9099-4341}, J.W.~Andrejkovic, T.~Bergauer\cmsorcid{0000-0002-5786-0293}, S.~Chatterjee\cmsorcid{0000-0003-2660-0349}, K.~Damanakis\cmsorcid{0000-0001-5389-2872}, M.~Dragicevic\cmsorcid{0000-0003-1967-6783}, A.~Escalante~Del~Valle\cmsorcid{0000-0002-9702-6359}, P.S.~Hussain\cmsorcid{0000-0002-4825-5278}, M.~Jeitler\cmsAuthorMark{2}\cmsorcid{0000-0002-5141-9560}, N.~Krammer\cmsorcid{0000-0002-0548-0985}, L.~Lechner\cmsorcid{0000-0002-3065-1141}, D.~Liko\cmsorcid{0000-0002-3380-473X}, I.~Mikulec\cmsorcid{0000-0003-0385-2746}, P.~Paulitsch, F.M.~Pitters, J.~Schieck\cmsAuthorMark{2}\cmsorcid{0000-0002-1058-8093}, R.~Sch\"{o}fbeck\cmsorcid{0000-0002-2332-8784}, D.~Schwarz\cmsorcid{0000-0002-3821-7331}, S.~Templ\cmsorcid{0000-0003-3137-5692}, W.~Waltenberger\cmsorcid{0000-0002-6215-7228}, C.-E.~Wulz\cmsAuthorMark{2}\cmsorcid{0000-0001-9226-5812}
% ...
\par}
\vskip\cmsinstskip
\dag:~Deceased\\
$^{1}$Also at Yerevan State University, Yerevan, Armenia\\
$^{2}$Also at TU Wien, Vienna, Austria\\
% ...
\end{sloppypar}
\end{document}